\documentclass[12pt]{iopart}

\usepackage{etoolbox}
\pdfoutput=1
\usepackage{lmodern}
\usepackage[T2A,T1]{fontenc}
\usepackage[utf8]{inputenc}
\usepackage{bm}
\usepackage{graphicx}

\usepackage{amssymb}
\usepackage{color}
\usepackage{xifthen}
\usepackage{shuffle}
\usepackage{multirow}
\usepackage{subcaption}
\usepackage[titletoc]{appendix}
\usepackage{cancel}
\usepackage{hyperref}
\usepackage{textcomp}
\usepackage{array}  

\usepackage{afterpage}
\usepackage{longtable}
\usepackage{caption}
\usepackage{calc}
\def\@oddfoot{\footnotesize\itshape {}}

\expandafter\let\csname equation*\endcsname\relax

\expandafter\let\csname endequation*\endcsname\relax
\usepackage{amsmath}

\newlength\imheight
\newlength\imwidth	 
\def\diagsizetwo{0.15\textwidth} 
\def\diagsize{0.19\textwidth} 

\newcommand{\num}[1]{
	\settoheight\imheight{\includegraphics{#1}}
	\settowidth\imwidth{\includegraphics{#1}}
	c \ifthenelse{ \lengthtest{\imheight < 0.9 \imwidth}}
	{  \left( 
		\parbox{\widthof{\scalebox{.96}{\includegraphics[width=\diagsize]{#1}}}}{\includegraphics[width=\diagsize]{#1}}
	}
	{ \left(		\parbox{\widthof{\scalebox{.96}{\includegraphics[height=\diagsizetwo]{#1}}}}{\includegraphics[height=\diagsizetwo]{#1}}		
	} 
	\right)}

\newcommand{\col}[1]{
	\settoheight\imheight{\includegraphics{#1}}
	\settowidth\imwidth{\includegraphics{#1}}
	c \ifthenelse{ \lengthtest{\imheight < 0.9 \imwidth}}
	{  \left( 
		\parbox{\widthof{\scalebox{.96}{\includegraphics[width=\diagsize]{#1}}}}{\includegraphics[width=\diagsize]{#1}}
		}
	{ \left(		\parbox{\widthof{\scalebox{.96}{\includegraphics[height=\diagsizetwo]{#1}}}}{\includegraphics[height=\diagsizetwo]{#1}}	
	} 
	\right)}

\def\P{{\rm P}}

\def\NeqFour{{{\cal N} = 4}}

\def\tree{{\rm tree}}

\def\pol{\varepsilon}

\def\calAthree{{\cal A}^{\tree}_3}
\def\calAfour{{\cal A}^{\tree}_4}

\def\cN{{\mathcal N}}

\def\f{\tilde f}
\def\T{T}

\def\calMthree{{\cal M}^{\tree}_3}
\def\calMfour{{\cal M}^{\tree}_4}
\def\calMfive{{\cal M}^{\tree}_5}
\def\calMsix{{\cal M}^{\tree}_6}

\def\Afive{A^{\tree}_5}

\def\be{\begin{equation}}
\def\ee{\end{equation}}
\def\eea{\end{eqnarray}}
\def\bea{\begin{eqnarray}}
\def\nn{\nonumber}

\def\sect#1{Sec.~{\ref{#1}}}

\def\fig#1{Fig.~{\ref{#1}}}

\def\eqn#1{Eq.~(\ref{#1})}

\def\eqns#1#2{Eqs.~(\ref{#1}) and~(\ref{#2})}

\def\spa#1.#2{\left\langle#1\,#2\right\rangle}
\def\spb#1.#2{\left[#1\,#2\right]}
\def\spash#1.#2{\spa{\smash{#1}}.{\smash{#2}}}
\def\spbsh#1.#2{\spb{\smash{#1}}.{\smash{#2}}}
\def\sand#1.#2.#3{%
\left\langle\smash{#1}{\vphantom1}^{-}\right|{#2}%
\left|\smash{#3}{\vphantom1}^{-}\right\rangle}
\def\sandpp#1.#2.#3{%
\left\langle\smash{#1}{\vphantom1}^{+}\right|{#2}%
\left|\smash{#3}{\vphantom1}^{+}\right\rangle}
\def\sandpm#1.#2.#3{%
\left\langle\smash{#1}{\vphantom1}^{+}\right|{#2}%
\left|\smash{#3}{\vphantom1}^{-}\right\rangle}
\def\sandmp#1.#2.#3{%
\left\langle\smash{#1}{\vphantom1}^{-}\right|{#2}%
\left|\smash{#3}{\vphantom1}^{+}\right\rangle}
\def\sand#1.#2.#3{%
  \left\langle\smash{#1}{\vphantom1}\right|{#2}%
  \left|\smash{#3}{\vphantom1}\right\rangle}
\def\sandp#1.#2.#3{%
  \left\langle\smash{#1}{\vphantom1}^{-}\right|{#2}%
  \left|\smash{#3}{\vphantom1}^{+}\right\rangle}
\def\sandpp#1.#2.#3{%
  \left\langle\smash{#1}{\vphantom1}^{+}\right|{#2}%
  \left|\smash{#3}{\vphantom1}^{+}\right\rangle}
\def\sandmm#1.#2.#3{%
  \left\langle\smash{#1}{\vphantom1}^{-}\right|{#2}%
  \left|\smash{#3}{\vphantom1}^{-}\right\rangle}
\def\sandpm#1.#2.#3{%
  \left\langle\smash{#1}{\vphantom1}^{+}\right|{#2}%
  \left|\smash{#3}{\vphantom1}^{-}\right\rangle}
\def\sandmp#1.#2.#3{%
  \left\langle\smash{#1}{\vphantom1}^{-}\right|{#2}%
  \left|\smash{#3}{\vphantom1}^{+}\right\rangle}

\DeclareMathAlphabet\mathbfcal{OMS}{cmsy}{b}{n}

\renewcommand\eqref[1]{\ref{#1}}

\newbox\charbox
\newbox\slabox
\def\s#1{{      
        \setbox\charbox=\hbox{$#1$}
        \setbox\slabox=\hbox{$/$}
        \dimen\charbox=\ht\slabox
        \advance\dimen\charbox by -\dp\slabox
        \advance\dimen\charbox by -\ht\charbox
        \advance\dimen\charbox by \dp\charbox
        \divide\dimen\charbox by 2
        \raise-\dimen\charbox\hbox to \wd\charbox{\hss/\hss}
        \llap{$#1$} }}

\usepackage{afterpage}
\usepackage{longtable}

\newcounter{homework}[section]  

\usepackage{etoolbox}

\makeatletter
\def\@mkboth#1#2{}
\newlength\appendixwidth
\preto\appendix{\addtocontents{toc}{\protect\patchl@section}}
\newcommand{\patchl@section}{%
  \settowidth{\appendixwidth}{\textbf{Appendix }}%
  \addtolength{\appendixwidth}{1.5em}%
  \patchcmd{\l@section}{1.5em}{\appendixwidth}{}{\ddt}%
}
\makeatother

\usepackage{cite}
\begin{document}

{\hfill SAGEX-22-03, NORDITA 2022-019, UUITP-18/22}
\vskip -1 cm 

\title[An Invitation to Color-Kinematics Duality and the Double Copy]{The SAGEX Review on Scattering Amplitudes \\ 
  Chapter 2: An Invitation to Color-Kinematics Duality and the Double Copy
 }

\author{Zvi Bern$^a$, John Joseph Carrasco$^{b,c}$, Marco Chiodaroli$^d$, \\ Henrik Johansson$^{d, e}$, Radu Roiban$^f$}

\address{$^a$ Mani L. Bhaumik Institute for Theoretical Physics, \\
~$\,$ Department of Physics and Astronomy, UCLA, Los Angeles, CA 90095, USA}
\address{$^b$ Department of Physics and Astronomy, 
\\
~$\,$ 
Northwestern University, Evanston, IL 60208, USA}
\address{$^c$ Institut de Physique Th\'{e}orique, Universit\'{e} Paris Saclay, \\
~$\,$ 
CEA, CNRS, F-91191 Gif-sur-Yvette, France}

\address{$^d$ Department of Physics and Astronomy, 
\\
~$\,$ 
Uppsala University, Box 516, 75120 Uppsala, Sweden}
\address{$^e$ Nordita, Stockholm University and KTH Royal Institute of Technology,\\  
~$\,$
Hannes Alfv\'{e}ns v\"{a}g 12, 10691 Stockholm, Sweden}
\address{$^f$ Institute for Gravitation and the Cosmos, \\
~$\,$ Pennsylvania State University, University Park, PA 16802, USA}

\ead{bern@physics.ucla.edu, carrasco@northwestern.edu, marco.chiodaroli@physics.uu.se, henrik.johansson@physics.uu.se, radu@phys.psu.edu}

\vspace{10pt}

\begin{abstract}
Advances in scattering amplitudes have exposed previously-hidden
color-kinematics and double-copy structures in theories ranging from gauge and
gravity theories to effective field theories such as chiral
perturbation theory and the Born-Infeld model.  These novel structures
both simplify higher-order calculations and pose tantalizing questions
related to a unified framework underlying relativistic quantum
theories. This introductory mini-review article invites further
exploration of these topics. After a brief introduction to
color-kinematics duality and the double copy as they emerge at tree and
loop-level in gauge and gravity theories, we present two distinct
examples: 1) an introduction to the web of double-copy-constructible theories, and 2)
a discussion of the application of the double copy to calculation relevant to
gravitational-wave physics. 
\end{abstract}

\maketitle

\tableofcontents

\section{Introduction}
\label{IntroductionSection}

Gauge and gravity theories share many formal similarities even though
their physical properties are distinct. Three of the known forces are
described by gauge theories and give interactions between
elementary particles, while gravity is a much weaker force that shapes
the macroscopic evolution of the universe and spacetime itself.
Nevertheless, the double-copy framework for gravity, which we outline
in this chapter, exploits a direct connection between these two classes of
theories, remarkably obtaining gravity directly from gauge theory.
This framework provides a fresh perspective on gravity and its
connection to the other forces, as well as very effective tools 
in the context of perturbative computations for gravity.  A more
comprehensive review may be found in Ref.~\cite{BCCJRReview}.

Modern ideas make it much easier to calculate scattering amplitudes in
perturbative quantum gravity compared to using Feynman rules.  When
one considers complete gauge-invariant scattering amplitudes instead
of individual Feynman diagrams, which are not gauge invariant, it
becomes possible to identify nontrivial structures.  The double copy
and the associated duality between color and
kinematics~\cite{KLT,BCJ,BCJLoop} are perhaps the most remarkable of
these structures, telling us that flat-space gravity scattering
amplitudes can be obtained directly from gauge-theory ones.  Via the
unitarity method~\cite{UnitarityMethod, Fusing, BDDPR,
  TripleCuteeJets, BCFUnitarity,FiveLoop} these same ideas can be
carried to loop level.  The double copy is central to our ability to
carry out calculations to very high loop orders in supergravity
theories in Minkowski vacua and a property of all supergravities whose
amplitudes have been analyzed in detail.  This leads to the natural
question on whether all (super)gravity theories are double copies of
suitably-chosen matter-coupled gauge theories.  The double copy offers
a possible unification of gauge and gravity theories in the sense of
providing a framework where calculations in both theories can be
carried out using the same building blocks, emphasizing that the two
types of theories are part of the same over-arching structure.  Beyond
gauge and gravity theories, double-copy relations also provide a new
perspective on quantum field theories, resulting in a web of theories,
linked by the same underlying building blocks (see
\sect{ZoologySection1} of this review and
e.g. Refs.~\cite{Cachazo2014xea, CheungUnifyingRelations, Chen2013fya,
  Du2016tbc, Chen2014dfa, Chen2016zwe, Cachazo2016njl,
  Carrasco2016ygv, He2016mzd, Elvang:2018dco, Cheung:2016prv,
  Cheung:2017yef, Chiodaroli2017ngp}).

The double copy has it origins in string theory.  In the 1980s, Kawai,
Lewellen, and Tye (KLT)~\cite{KLT} realized that open- and
closed-string tree-level amplitudes both share the same fundamental
gauge-invariant kinematic building blocks.  They showed that
closed-string tree amplitudes could be written as a sum over products
of pairs of open-string tree amplitudes.  In the low-energy limit,
this translates directly to relations between gauge and gravity
field-theory amplitudes for any number of external
particles~\cite{MultiLegOneLoopGravity}.  The double copy is
streamlined and systematized by the introduction of the duality
between color and kinematics~\cite{BCJ}.  The duality effectively
states that scattering amplitudes in gauge theories---and, more
generally, in theories with some continuous internal symmetry
algebra---can be rearranged so that kinematic building blocks obey the
same generic algebraic relations as their color factors.  Via the
duality, not only can we constrain the kinematic dependence of each
graph, but we can also convert gauge-theory scattering
amplitudes to gravity ones.  This is done through the simple replacement: color
$\Rightarrow$ kinematics. Such constructions have been summarized by
the heuristic statement ``gravity $\sim$ (gauge theory) $\times$
(gauge theory)''.

At tree level, proofs exist~\cite{BjerrumMomKernel,
  MafraExplicitBCJNumerators, BjerrumManifestingBCJ, DuTengBCJ,
  delaCruz:2017zqr, Bridges:2019siz} that the duality and double copy
hold.  At loop level, less is known, but explicit constructions
show that the duality between color and
kinematics and the double copy hold for a wide class of examples~\cite{BCJLoop, Neq44np, FivePointN4BCJ, WhiteIRBCJ,
  SimplifyingBCJ, Du:2012mt, Yuan:2012rg, FourLoopFormFactor,
  Boels:2013bi, Bjerrum-Bohr:2013iza, OneTwoLoopPureYMBCJ,
  Ochirov:2013xba, MafraSchlottererTwoLoop, BCJDifficulty,
  HeMonteiroSchlottererBCJNumer, FiveLoopFormFactor, BoelsFourLoop,
  HeSchlottererZhangOneLoopBCJ, JohanssonTwoLoopSusyQCD,
  Jurado:2017xut, Boels:2017ftb, Faller:2018vdz}.
A natural question is whether the double copy carries over to
classical solutions beyond scattering amplitudes, especially for
gravity.  Scattering amplitudes in flat space are gauge invariant and
independent of coordinate choices, while generic classical solutions do
depend on such choices, complicating the problem of relating gravity
solutions to gauge-theory ones.
Nevertheless, there has been substantial progress in unraveling both
the underlying principles of color-kinematics duality~\cite{Square,
WeinzierlBCJLagrangian, Monteiro2011pc, OConnellAlgebras,
Monteiro:2013rya, Ho:2015bia, Fu:2016plh, BjerrumManifestingBCJ,
Fu:2018hpu,Chen:2019ywi, Cheung:2017yef, Brandhuber:2021bsf,
Ben-Shahar:2021zww} and finding explicit examples of classical
solutions related by the double-copy property~\cite{Saotome:2012vy,
Monteiro2014cda, Luna2015paa, Ridgway2015fdl, Luna2016due,
White2016jzc, Cardoso2016amd, Goldberger2016iau, Luna2016hge,
Goldberger2017frp, Adamo2017nia, DeSmet2017rve, BahjatAbbas2017htu,
CarrilloGonzalez2017iyj, Goldberger2017ogt, Li2018qap,
Ilderton:2018lsf, Lee:2018gxc, Plefka:2018dpa, ShenWorldLine,
Berman:2018hwd, Gurses:2018ckx, Adamo:2018mpq, Bahjat-Abbas:2018vgo,
Luna:2018dpt, Farrow:2018yni, CarrilloGonzalez:2019gof, PV:2019uuv}.
One of the most striking applications of the double copy beyond
scattering amplitudes relates to gravitational-wave physics, as
highlighted by Refs.~\cite{Goldberger2016iau, ShenWorldLine, CheungPM,
Kosower:2018adc, 3PM, Buananno3PMCheck,3PMLong}.

This short review is organized as follows.  In \sect{ColorKinematics}
we give an overview of color-kinematics duality and the associated double copy
for the simplest case of pure gauge theory.  Then
in \sect{ZoologySection1} we summarize the status of the web of
theories linked by double copy relations, for both gravitational and
non-gravitational theories.  Then in \sect{ClassicalDoubleCopySection}
we describe the application of the double copy to the problem
gravitational-wave physics.  Some brief comments on the outlook are given 
in \sect{ConclusionSection}.

\section{Color/kinematics duality and the double copy}
\label{ColorKinematics}

\subsection{Basics of color/kinematics duality}

The canonical example of a theory exhibiting color-kinematics duality is a gauge theory in which all fields are in 
the adjoint representation of the gauge group, as considered in the original paper~\cite{BCJ}.
In any such theory, the $m$-point tree-level amplitudes in $D$ dimensions may be written as
\begin{equation}
i\mathcal{A}^{\tree }_{m} = g^{m-2} \sum_{j}
\frac{c_{j} n_{j}}{\prod_{i_{j}} d_{i_{j}}} \,,
\label{CubicRepresentation}
\end{equation}
where the sum runs over the set of distinct $m$-point graphs with only
three-point vertices. Contributions from any
diagram with quartic or higher-point vertices can be assigned to these graphs
simply by multiplying and dividing by appropriate missing propagators.
The color factor $c_j$ is obtained by dressing each vertex in graph
$j$ with the relevant group-theory structure constant,
$\tilde{f}^{abc}=i\sqrt{2}f^{abc} = \mathrm{Tr}([T^{a},T^{b}] T^{c})$,
where the hermitian generators of the gauge group $T^a$ are normalized as
$\mathrm{Tr}(T^{a}T^{b})=\delta^{a b}.$ 
The kinematic numerators $n_{j}$ depend on momenta, polarizations,
and spinors, as one would obtain using Feynman rules.
The factors $1/d_{i_{j}}$ are ordinary scalar Feynman propagators,
where $i_j$ runs over the propagators for diagram $j$.
We denote the gauge-theory coupling constant as $g$. 

\begin{figure}
\begin{center}
\includegraphics[scale=.42]{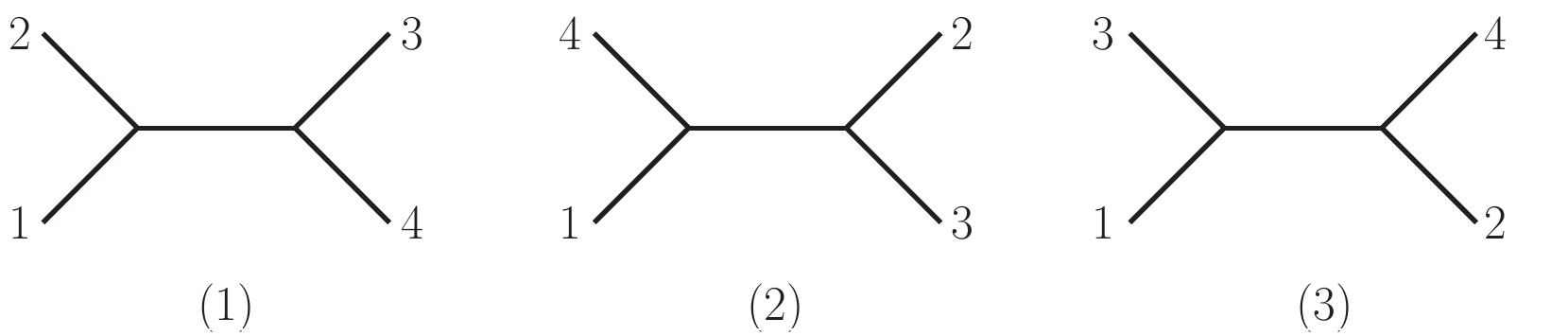}
\end{center}
\caption{The three diagrams with cubic vertices describing a four-point tree amplitude.}
\label{FourPointFigure}
\end{figure}

The nontrivial insight is that the kinematic numerators can be made to
obey the same algebraic relations as the color
factors~\cite{BCJ,BCJLoop,SimplifyingBCJ, BCCJRReview}.  For theories
with only fields in the adjoint representation there are two generic
properties.  
The first is that they obey Jacobi relations that are inherited from the Lie algebra structure.
For example, for the diagrams in \fig{FourPointFigure} the color factors obey
\begin{align}
f^{a_1 a_2 b} f^{b a_3 a_4}+f^{a_1a_4bi} f^{b a_2 a_3} + f^{a_1 a_3 b} f^{b a_4 a_2} = 0 \,.
\label{ColorJacobiD}
\end{align}
Such Lie-algebra relations are directly tied to the gauge
invariance of amplitudes. For each color Jacobi identity we then demand that there be a 
corresponding identity for the kinematic numerators,
\begin{equation}
c_{i}+c_{j}+c_{k} = 0\ \qquad \Rightarrow\  \qquad n_{i}+n_{j}+n_{k} = 0 \,,
\label{BCJDuality}
\end{equation}
where $i$, $j$, and $k$ refer to three graphs which are identical
except for one internal edge.  
  A second property is that kinematic factors should
have the same antisymmetry under twists of diagrams lines as color
factors
\begin{equation}
 c_{\overline{i}} = - c_{i} \qquad \Rightarrow \qquad  n_{\overline{i}}=- n_{i}\, ,
\label{BCJFlipSymmetry}
\end{equation}
where the graph $\overline{i}$ is graph $i$ with twisted lines.
For example, the color factor of diagram 1 of \fig{FourPointFigure} 
is antisymmetric under the swap of legs 1 and 2; we then require 
the corresponding kinematic numerator exhibits the same antisymmetry.

The algebraic properties of color factors in gauge-theory
amplitudes have important implications for kinematic numerators
in \eqn{CubicRepresentation}.  Consider a gauge-theory amplitude where
we shift the numerators,
\begin{equation}
 n_i = n_i'- \Delta_i  \, ,
\label{Shift}
\end{equation}
subject to the constraint,
\begin{equation}
\sum_{i}\, \frac{c_{i} \Delta_i}{D_i}  = 0 \, ,
\label{GenGauage}
\end{equation}
the amplitude (\ref{CubicRepresentation}) is unchanged.
Given that color factors are not independent but satisfy linear relations, nontrivial shifts of the kinematic
numerators that leave the amplitudes invariant can be found. 
The $\Delta_i$ can be thought of as generalized gauge functions that drop out of the amplitude.

When we have numerators $n_i$ that obey the same algebraic relations
as the color factors $c_i$  in \eqns{BCJDuality}{BCJFlipSymmetry},
we can then replace 
\begin{equation} c_i \rightarrow n_i\,,
\label{ColorSubstitution}
\end{equation}
in any given formula or amplitude. Given that the algebraic properties of
the kinematic numerators are the same as those of the color factors, the
new amplitude that results will also satisfy a generalized
gauge invariance.  Remarkably, this color-to-kinematics
replacement gives us gravity amplitudes,
\begin{equation}
i \mathcal{M}^{\tree}_{m} =  \left(\frac{\kappa}{2}\right)^{m-2}
\sum_{j}\frac{\tilde{n}_{j}n_{j}}{D_j} \,,
\label{DoubleCopy}
\end{equation}
where $\kappa^2 = 32 \pi G$ with $G$ Newton's constant, and where
$\tilde{n}_j$ and $n_j$ are the kinematic numerator factors of
the two gauge-theory amplitudes.   (Note that $\mathcal{M}$ follows the usual Feynman diagram normalization so that the scattering matrix is $S=1+i T = 1+\mathcal{M}$.)
The two gauge theories can be different.
Only one of the two sets of numerators needs to manifestly satisfy the
duality~(\ref{BCJDuality}) for the double-copy~(\ref{DoubleCopy}) to be 
gauge-invariant~\cite{BCJLoop,Square}.

Similar properties are conjectured to hold at loop level.
Analogous to the tree level case (\ref{CubicRepresentation}),
an $L$-loop $m$-point gauge theory scattering amplitude can then be organized as,
\begin{equation}
{\cal A}^{(L)}_{m} = i^{L-1} g^{m-2+2L} \sum_{i}\, 
 \int \prod_{l = 1}^{L} \frac{d^D\ell_l}{(2\pi)^{D}} \frac{1}{S_i} \frac{c_{i} n_i}{\prod_{i_{j}} d_{i_{j}}} \, ,
 \label{gaugeAmp}
 \end{equation}
where the sum runs over the distinct $L$-loop $m$-point diagrams with
only cubic vertices.  Each such diagram corresponds to a unique color
factor $c_i$.\footnote{Our conventions for the overall phase
  in the representations of gauge-theory and gravity amplitudes follow
  the one in Ref.~\cite{Chiodaroli2017ngp} rather than the original
  Bern--Carrasco--Johansson (BCJ) papers \cite{BCJ,BCJLoop}.}
It also has an associated
denominator corresponding to the product of the denominators of the
Feynman propagators $\sim 1/d_{i_{j}}$ of each internal line of the
  diagram. A difference with tree level is that one needs to include
 symmetry factors  $S_i$ that remove internal overcount of loop
  diagrams; they can be computed, as for regular Feynman diagrams, by counting the number of discrete
  symmetries of each diagram with fixed external legs.  
As for tree level, the representation of the amplitude in terms of
cubic diagrams is trivial.  The nontrivial part is to find
representations of the amplitude where the duality holds so that the
integrand kinematic numerators $n_i$ satisfy the duality in
\eqns{BCJDuality}{BCJFlipSymmetry}.   Whether
this can be done in general at loop level remains a conjecture, although there is
considerable evidence that such representations can be
found~\cite{BCJLoop, Neq44np, FivePointN4BCJ, WhiteIRBCJ,
  SimplifyingBCJ, Du:2012mt, Yuan:2012rg, FourLoopFormFactor,
  Boels:2013bi, Bjerrum-Bohr:2013iza, OneTwoLoopPureYMBCJ,
  Ochirov:2013xba, MafraSchlottererTwoLoop, BCJDifficulty,
  HeMonteiroSchlottererBCJNumer, FiveLoopFormFactor, BoelsFourLoop,
  HeSchlottererZhangOneLoopBCJ, JohanssonTwoLoopSusyQCD,
  Jurado:2017xut, Boels:2017ftb, Faller:2018vdz}.  
However, in certain
cases, such as the five-loop four-point amplitude of ${\mathcal N} =
4$ super-YM theory,  such representations have been elusive.
In other cases, such as the all-plus two-loop five-gluon amplitude in
pure-YM theory, the BCJ form of the amplitude has a superficial
power-count much worse than that of standard Feynman
diagrams~\cite{BCJDifficulty} leading to more complicated
expressions.

Consider two $m$-point $L$-loop gauge theory amplitudes, ${\cal A}^{(L)}_{m}$ and ${\widetilde {\cal
    A}}^{(L)}_{m}$, and assume that they are organized as in~\eqn{gaugeAmp}. 
 Furthermore, label the two sets of numerators for each amplitude $n_i$ and
$\tilde n_i$, respectively. If at least one of the amplitudes, say
${\widetilde {\cal A}}^{(L)}_{m}$, manifests the duality, we may now replace the color factors of the first amplitude
with the duality-satisfying numerators $\tilde n_i$ of the second
one. This gives the loop-level double-copy formula for gravitational scattering
amplitudes \cite{BCJ,BCJLoop},
\begin{equation}
{\cal M}^{(L)}_{m} =    {\cal A}^{(L)}_{m} \Big|_{c_i \! \rightarrow  \tilde n_i 
                            \atop g\rightarrow  {\kappa}/{2}}
  =  i^{L-1}\;\!\Big(\frac{\kappa}{2}\Big)^{m-2+2L} \sum_{i}\, \int \prod_{l = 1}^L\frac{d^{D}\ell_l}{(2\pi)^{D}} \frac{1}{S_i} \frac{n_i \tilde{n}_i}{D_i} \,,
\label{DCformula}
\end{equation}
where the gravitational coupling $\kappa/2$ compensates for the
change of engineering dimension when replacing color factors with
kinematic numerators.  The most challenging aspect of double-copy construction is finding a
representation of the gauge-theory integrand that satisfies the duality
in \eqns{BCJDuality}{BCJFlipSymmetry}.  For the replacement
(\ref{ColorSubstitution}) to be valid under the integration symbol, it
is important that the color factors not be explicitly evaluated by
summing over the contracted indices. Under explicit evaluation it can
turn out that certain color factors vanish, either by antisymmetry or
by a special property of the group under consideration.  We do not
wish to impose any specific color-factor properties on the numerator
factors, only generic ones.

Standard methods such as
Feynman rules, on-shell recursion~\cite{BCFW}, or generalized
unitarity~\cite{UnitarityMethod,Fusing,TripleCuteeJets,BCFUnitarity},
generally do not naturally result in  numerators obeying the duality.  One
straightforward (albeit somewhat tedious) way to find such numerators is
to use an Ansatz which is constrained to manifest the duality and
to match the correct amplitude~\cite{FivePointN4BCJ,
SimplifyingBCJ}.  Constructive ways to obtain numerators also
exist~\cite{Mafra:2009bz, BjerrumMomKernel,
MafraExplicitBCJNumerators, Mafra:2014oia,
BjerrumManifestingBCJ,Carrasco2016ldy, DuTengBCJ, delaCruz:2017zqr,
Bridges:2019siz}.

Aside from amplitudes, the duality has also been demonstrated to hold for
currents with one off-shell leg~\cite{FourLoopFormFactor,
FiveLoopFormFactor, PureSpinorsBCJAmplProof, Mafra:2016ltu,
PureSpinorsBCJAmplProof, Mafra2016mcc,
SchlottererBGCurrent,Jurado:2017xut, Boels:2017ftb}.  A possible
way to make the duality valid for general off-shell quantities would be
to find a Lagrangian that generates Feynman rules whose diagrams
automatically respect the duality. Such  Lagrangians are known to a
few orders in perturbation
theory~\cite{Square,WeinzierlBCJLagrangian,Vaman:2014iwa,
Mastrolia:2015maa}. An important problem is to find a useful closed form of
such a Lagrangian valid to all orders.

\subsection{Gauge-theory amplitude relations}

The duality also implies that there are nontrivial relations between
partial amplitudes, which are gauge invariant subdivisions of gauge
theory scattering amplitudes.  At tree level, with all particles in
the adjoint representation of $SU(N_c)$, a full tree amplitude can be
decomposed into partial amplitudes,
\begin{equation}
{\cal A}^\tree_n (1,2,3, \ldots, n)=g^{n-2}
\sum_{{\rm noncyclic}} {\rm Tr}[T^{a_1}T^{a_2} T^{a_3}\cdots T^{a_n}]
\, A^\tree_n (1,2,3, \ldots, n)\,,
\label{TreeDecomposition}
\end{equation}
where $A_n^\tree$ is a tree-level color-ordered $n$-point partial
amplitude.  The sum is over all noncyclic
permutations of legs, which is equivalent to all permutations keeping
leg $1$ fixed.  Helicities and polarizations are suppressed.
Reviews of such color decompositions are found in Refs.~\cite{ManganoParkeReview, TasiLance,
BDKUniarityReview, ElvangHuangReview}.

The generalized gauge invariance (\ref{GenGauage}) has an interesting
consequence:  it leads to nontrivial relations between gauge-theory partial amplitudes,
known as BCJ amplitude relations, 
\begin{align}
s_{24} A_4^\tree(1,2, 4, 3) &=  s_{14} A_4^\tree(1, 2, 3, 4) \,,
\nn\\
s_{24} A_5^\tree(1,2, 4, 3, 5) &=
 (s_{14} + s_{45}) A_5^\tree(1,2, 3, 4,5) +
  s_{14}  A_5^\tree(1, 2, 3, 5, 4)  \,,\phantom{xxxxxxxx}
\nn\\ 
s_{24} A_6^\tree(1, 2, 4, 3, 5, 6) &=
(s_{14} + s_{46}+s_{45}) A_6^\tree(1, 2, 3, 4, 5, 6) \nn\\
& \null
+ (s_{14} + s_{46}) A_6^\tree(1, 2, 3, 5, 4, 6)
+  s_{14} A_6^\tree(1, 2, 3, 5, 6, 4) \,, 
\label{AmplitudeRelations}
\end{align}
At tree level such relations exist for any number of external legs~\cite{BCJ}.  Progress
at loop level has been more difficult, except for special kinematic configurations such the forward limit~\cite{PierreHigherLoopMonodromy, Hohenegger:2017kqy, HeSchlottererZhangOneLoopBCJ, PierreOneLoopMonodromy, Chiodaroli2017ngp, Tourkine:2019ukp}.

\subsection{KLT formula and constructive tree-level adjoint color-kinematics duality  \label{secKLT}}

\newcommand\eedot[2]{(\varepsilon_{#1}\cdot\varepsilon_{#2})}
\newcommand\ke[2]{(k_{#1}\cdot\varepsilon_{#2})}

Double-copy relations have been known since 1985 in the form of Kawai-Lewellen-Tye relations~\cite{KLT}.
We now review these relations from the vantage point of color-kinematics duality.
At three points, the full color-dressed amplitude for Yang-Mills in $D$  dimensions is simply given by
\begin{equation}
      i  \calAthree = g \f^{a_1a_2a_3} n_{123}\,,
\end{equation}
where $g$ is the gauge-theory coupling constant, $\f^{a_1a_2a_3}$ is the
suitably-normalized color structure constant for the gauge theory,
and  $n_{123}$  is the on-shell Feynman three-vertex,
\begin{equation}
  n_{123}= \sqrt{2}\left( \eedot{1}{2}\, \ke{2}{3} + \eedot{2}{3} \, \ke{3}{1} - \eedot{1}{3}\, \ke{3}{2} \right)\,.
  \label{threegluons}
\end{equation}
The $k_i$ and $\varepsilon_j$ are the momenta and polarizations of the external legs.  We can 
think of $n_{123}$ as the kinematic numerators described above, although here there is no propagator denominators.
It is straightforward to see that this is fully antisymmetric under exchange between any pair of leg labels.
As this satisfies the duality between color and kinematics it can be immediately be used in the construction of a three-point gravitational amplitude,
\begin{equation}
   i \calMthree= \left(\frac{\kappa}{2}\right)  n_{123} \, \widetilde{n}_{123} \,.
  \label{doublecopy3}
\end{equation}
where $\kappa/2$ is the gravitational coupling.  Note that, in the case
of three points, there is no gauge freedom.  The $n_{123}$ can be
interpreted as gauge-theory ordered (``color stripped'') amplitudes
and we see the simplest example of the tree-level KLT relations
between ordered gauge-theory amplitudes and tree-level gravitational
amplitudes,
\begin{equation}
 -i \calMthree(1,2,3) = \left(\frac{\kappa}{2}\right)
A_3^\tree(1,2,3) \widetilde A_3^\tree(1,2,3)\,. 
\label{threepointdoublecopy}
\end{equation}

The situation is more interesting at four-points.  Here we have the
freedom to arrive at different representations for each of the three
distinct labelings $n_s, n_t, n_u$ of the cubic graphs labeled by the
Mandelstam invariant describing each graph's propagator, $s=(k_1+k_2)^2$, $t=(k_2+k_3)^2$, and $u=-s-t$.  The four-point
amplitude is simply
\begin{equation}
  i\calAfour= g^2 \left(\frac{n_s c_s}{s} + \frac{n_t c_t}{t} + \frac{n_u c_u}{u}\right).
  \label{fourGaugeAmp}
\end{equation}
corresponding to $m=4$ in \eqn{CubicRepresentation}.

We can decompose the amplitude (\ref{fourGaugeAmp}) into color-ordered partial amplitudes using \eqn{TreeDecomposition}, 
which is expressed in terms of the kinematic numerators, 
\begin{align}
	\label{fourPntAmps}
	iA_{st} &\equiv iA^\tree_4(1,2,3,4) = \frac{n_s}{s} - \frac{n_t}{t} \,,  \\
	iA_{tu} & \equiv iA^\tree_4(1,3,2,4) =  \frac{n_t}{t}-\frac{n_u}{u}  \,,  \\
	iA_{us} & \equiv iA^\tree_4(1,2,4,3) =\frac{n_u}{u} - \frac{n_s}{s} \,,
\end{align}
where the signs follow from antisymmetry of color factors.
At first sight, it might seem that, with three kinematic numerators and three ordered
amplitudes, we might be able to invert this set of linear relations to express
the numerators in terms of amplitudes.
However, since kinematic numerators satisfy  $n_s+n_t+n_u=0$,
the matrix is singular and cannot be inverted.  Reducing the linear relations, one
finds that all of the ordered amplitudes are related by the BCJ
relations described earlier in \eqn{AmplitudeRelations}, which 
we can write in an equivalent permutation-invariant form as follows, 
\begin{equation}
\label{bcjFour}
 s t A_{st} = u t A_{tu} = s u A_{us} \,.
\end{equation}
Using \eqn{fourPntAmps}, we can solve $n_u$ in terms of $A_{st}$ and $n_t$,
\begin{equation}
\label{nuSolve}
n_u = -s (iA_{st})+ \frac{u}{t} n_t  \,.
\end{equation}
Remarkably, $n_t$ cancels out when we substitute $n_u$ into \eqn{fourPntAmps} and solve 
in terms of  $A_{st}$.   Indeed,
plugging \eqn{nuSolve} and $n_s=-(n_t+n_u)$  into \eqn{fourPntAmps} simply
produces the four-point BCJ amplitude relations (\ref{AmplitudeRelations}), and doing the same to \eqn{fourGaugeAmp},
yields the four-point amplitude in a basis of color factors,
\begin{align}
	 \calAfour & = g^2 \left(c_s A_{st} - c_u \frac{s}{u} A_{st} \right) 
	            = g^2 \left(c_s A_{st} - c_u A_{tu} \right) \, .
\end{align}

By applying the above equations expressing the numerators in terms of
$A_{st}$ and $n_t$ to the double copy in \eqn{DoubleCopy} with $m=4$, we can 
thereby obtain the gravitational amplitude in terms of ordered gauge-theory
amplitudes~\cite{BCJ},
\begin{align}
i \calMfour & = \left(\frac{\kappa}{2}\right)^2  \left(\frac{n_s \widetilde{n}_s}{s}+\frac{n_t \widetilde{n}_t}{t}+\frac{n_u \widetilde{n}_u}{u} \right) \nn \\
  & = \left(\frac{\kappa}{2}\right)^2 \left( s t A_{st} \right) \left( s t \widetilde A_{st} \right)  \left( s t u \right)^{-1} \nn \\
  & = \left(\frac{\kappa}{2}\right)^2 s A_{st}  \widetilde{A}_{su}\,,
  \label{KLT4}
\end{align}
where we used the BCJ amplitude relations (\ref{bcjFour}) to obtain the final form.
The relations (\ref{bcjFour}) allow us to find many equivalent ways of expressing the four-point KLT relations.
A similar exercise may be carried out at any multiplicity.  Sample relations through six points are
\begin{align}
 \calMfive =
  \null &  i\, \left(\frac{\kappa}{2}\right)^{3} \left(  s_{12} s_{45} \Afive(1,2,3,4,5)  \widetilde A_5^\tree(1,3,5,4,2)  \right.\nn\\
  & \hskip 1.2 cm
\left.   +  \, s_{14} s_{25}\Afive(1,4,3,2,5) \widetilde A_5^\tree(1,3,5,2,4) \right), \nn \\
 \calMsix = \null & -i \left(\frac{\kappa}{2}\right)^{4} \left(
s_{12} s_{45} A_6^\tree(1,2,3,4,5,6) \bigl(s_{35} \widetilde A_6^\tree(2,1,5,3,4,6) \right.\nn \\
&\hskip 1.2 cm  
\left.   + \, (s_{34} + s_{35} ) \widetilde A_6^\tree(2,1,5,4,3,6) \bigr)  + {\cal P}(2,3,4) \right) ,
\label{KLT456}
\end{align}
where ${\cal P}(i,j,k)$ represents a sum over all permutations of leg labels $i,j,k$.  These are exactly 
the low-energy limit of the KLT relations~\cite{KLT}.

These relations have an $m$-point generalization in terms of a basis of
$(m-3)! \times (m-3)!$ ordered gauge amplitudes~\cite{MultiLegOneLoopGravity}:
\begin{align}
{\cal M}_m^\tree = -i \left(\frac{\kappa}{2}\right)^{m-2}
\!\!\!\!\!\!\!\!\!  \sum_{\sigma,\rho \in S_{m-3}(2,\dots,m-2)}\!\!\!\!\!\! \!\!\!\!\!\! 
A_m^\tree(1,\sigma,m-1,m)S[\sigma|\rho]\widetilde{A}_m^\tree(1,\rho,m,m-1)\,.
\label{KLT}
\end{align}
The formula makes use of a matrix $S[\sigma|\rho]$ known as the field-theory KLT or momentum kernel.
This is an
$(m-3)! \times (m-3)!$ matrix of kinematic polynomials that acts on
the vector of $(m-3)!$  independent color-ordered amplitudes~\cite{KLT, MultiLegOneLoopGravity,BjerrumBohr2010ta,BjerrumBohr2010yc,BjerrumMomKernel}:
\begin{align}
 S[\sigma|\rho]=\prod_{i=2}^{m-2}\biggl[
2p_1\cdot p_{\sigma_i}+\sum_{j=2}^{i}2p_{\sigma_i}\cdot p_{\sigma_j}\theta(\sigma_j,\sigma_i)_\rho\biggr]\,,
 \label{momKernel}
\end{align}
where $\theta(\sigma_j,\sigma_i)_\rho=1$ if $\sigma_j$ is before $\sigma_i$ in the permutation $\rho$, and zero otherwise.  Compact recursive presentations of the KLT kernel have been found in Refs.~\cite{BjerrumMomKernel,Carrasco2016ldy}.

\begin{figure}[tb]
\begin{center}
\includegraphics[width=2.6in]{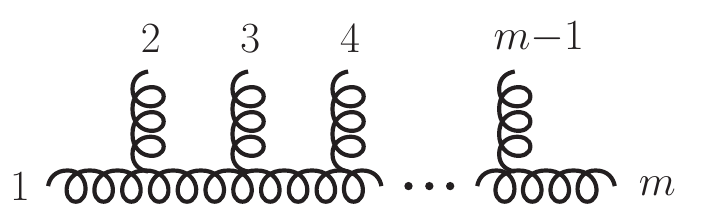}
\end{center}
\caption{An $m$-point half-ladder tree diagram.  
}
\label{HalfLadderFigure}
\end{figure}

There are a number of explicit constructions of the kinematic numerators that satisfy color-kinematics duality for arbitrary number of  
external particles.  The first of these was based on matching to KLT relations~\cite{KiermaierTalk,BjerrumMomKernel}
and making use of the  Del Duca-Dixon-Maltoni (DDM) color-basis~\cite{DixonMaltoni}.  The result
are kinematic numerators for the half-ladder (or multi-peripheral) diagrams,
as depicted in \fig{HalfLadderFigure}, with the all remaining numerators determined by kinematic Jacobi relations.
A valid specification for the half-ladder is given by the above KLT kernel,
\bea
n(1,\sigma(2,\dots, m-2), m-1,m) &=& -i \sum_{\rho\in S_{m-3}} S[\sigma | \rho] \widetilde A_m^{\tree}(1,\rho,m, m-1)\,, \nn \\
n(1,\tau(2,\dots, m-1),m)\Big|_{\tau(m-1) \ne m-1} \!\!\!\!\!\!\!\!\!\!\!\!\!\!\!\!&=& 0 \,.
 \label{nonlocalNum}
\eea All remaining $(2m-5)!!-(m-2)!$ numerators are determined by the
Jacobi relations.  Because the numerators are expressed in terms of
amplitudes which are nonlocal, this representation has the
disadvantage of resulting in nonlocal numerators.  It also does not give
manifestly crossing-symmetric results, although the generated amplitudes
do, of course, satisfy crossing.  One can find crossing-symmetric
kinematic numerators either by solving the Jacobi relations as
functional constraints via an Ansatz~\cite{virtuousTrees} or by
appropriately symmetrizing \eqn{nonlocalNum}, as in
Ref.~\cite{Naculich:2014rta}.   There are by now a number of efficient means of generating
arbitrary multiplicity tree-level Yang-Mills color-dual numerators with varying degrees of 
manifest crossing symmetry, see e.g.~Refs.~\cite{Edison:2020ehu,Cheung:2021zvb, Brandhuber:2021bsf} and references therein.

\subsection{Color-kinematics and double-copy construction beyond the adjoint representation \label{nonadjck}}

\def\tf{f}

\begin{figure}[t]
\centering
\includegraphics[scale=0.35,trim=0 0 0 0,clip=true]{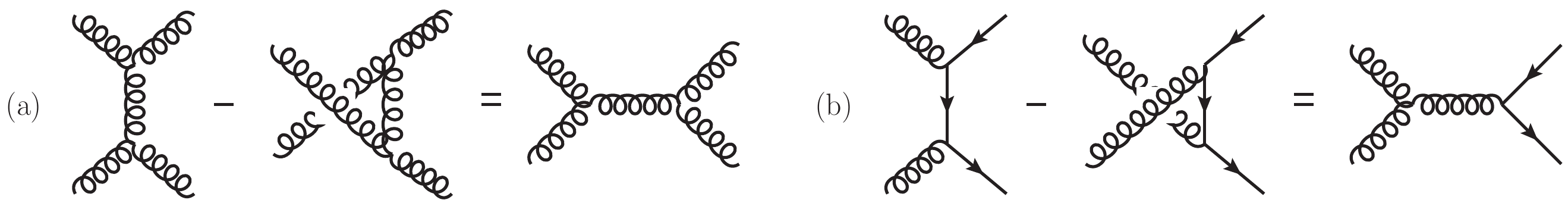}
\vspace{-3pt}
\caption{\small Graphical representation of the color-algebra relations in the adjoint~(a) and
  some arbitrary representation~(b).  The curly lines represent adjoint
  representation states and the straight lines the arbitrary
  representation. }
\label{fig:arbjacobi}
\end{figure}

As discussed above, amplitudes with adjoint fields can manifest the
duality between color and indeed lead naturally to supersymmetric
theories~\cite{Chiodaroli2013upa,Weinzierl:2014ava, BCCJRReview}.
What about matter fields in the fundamental or more general color
representations?
First we consider a gauge-theory with arbitrary
gauge group and with matter particles---spin 0 or spin
$\frac{1}{2}$---transforming in some matter representation of that
gauge group. For simplicity, we will restrict to cases where the only color tensors appearing in amplitudes are
$\f^{abc}$ and $(\T^a)_{i}^{ \ j}$ which both have three free
indices. Thus, all color factors can again correspond to cubic
diagrams and with appropriate normalization satisfy the defining
commutation relations,
\begin{align}
\f^{dae} \f^{ebc}- \f^{dbe} \f^{eac}&= \f^{abe} \f^{ecd}\,, \nn \\
(\T^a)_i^{~k}(\T^b)_k^{~j}-(\T^b)_i^{~k}(\T^a)_k^{~j}
  &=  \f^{abc} (\T^c)_i^{~j}\,,
\label{colorAlgRelations}
\end{align}
as depicted in \fig{fig:arbjacobi}. We find it convenient to introduce raised and lowered indices commonly associated with complex representations.

A difference with the pure-adjoint case is that edges of graphs now also encode the relevant representation, see e.g. \fig{fig:arbjacobi}. 
While important, many of the same ideas and approaches apply. We can still write $m$-point tree amplitudes in terms of cubic graphs,
\begin{equation}
   {\cal A}^{\text{tree}}_{m,k} = - i g^{m-2}
      \sum_{i} \frac{c_i n_i}{D_i} \,,
\label{BCJformYM}
\end{equation}
where $c_i$ are color factors, $n_i$ are kinematic numerators, and
$D_i$ are denominators encoding the propagator structure of the cubic diagrams.
The denominators (and numerators) may in principle contain masses,
corresponding to massive  propagators. 
The color factors $c_i$ in \eqn{BCJformYM} are constructed
from the cubic diagrams using two building blocks:
the structure constants $\f^{abc}$ for three-gluon vertices
and generators $(\T^a)_{i}^{ \ j}$ for quark-gluon vertices.
When separating color from kinematics,
the diagrammatic crossing symmetry only holds up to signs
dependent on the permutation of legs.
These signs are apparent in the total antisymmetry of $\f^{abc}$. For a uniform treatment of the generic representations, 
it convenient to  introduce a similar antisymmetry for the fundamental generators, artificially if necessary,
\begin{equation}
 (\T^a)_{\ i}^{j}  \equiv - (\T^a)_{i}^{\ j} ~~~~ \Leftrightarrow~~~~\f^{cab} = - \f^{bac} \,.
\label{signflip}
\end{equation}
This allows us to introduce a compatible  antisymmetry 
in color-ordered kinematic vertices, so that they are effectively the same as for the adjoint representation. 
As noted in \eqn{colorAlgRelations} the color factors obey Jacobi and commutation identities.
They both imply three-term color-algebraic relations of the form given in
\eqn{BCJDuality}.  The existence of algebraic relations between factors~$c_i$ means that the
corresponding kinematic coefficients~$n_i/D_i$  need not be
unique nor independently gauge-invariant.

As with the adjoint representation, one can still solve for all color factors
in terms of a minimal basis exploiting relevant antisymmetry and Jacobi-like identities, \eqn{colorAlgRelations}. 
Using this basis in the full amplitude allows the identification of  gauge-invariant ordered amplitudes as kinematic
coefficients of the remaining color weights. These gauge-invariant ordered amplitudes will be related to each other
by virtue of the fact that the kinematic weights $n_i$ can be arranged in a color-dual fashion.
A general color decomposition of tree-level amplitudes with matter representations may be found in
Ref.~\cite{Johansson:2015oia} (see also Refs.~\cite{Melia:2013bta,Melia:2015ika,Ochirov:2019mtf}). These ideas have been applied to massive scalar QCD at tree and loop level in Ref.~\cite{Carrasco:2020ywq} and to $\mathcal{N}=2$ super-QCD with $N_f$ fermionic hypermultiplets in the fundamental through two-loops in Ref.~\cite{Johansson:2017bfl, Duhr:2019ywc}.  Further discussions of massive theories are found in Refs.~\cite{Chiodaroli:2015rdg, Johnson:2020pny,
Moynihan:2020ejh,  Gonzalez:2021bes, Hang:2021oso, Li:2021yfk, Gonzalez:2022mpa}.

Consider now generic single color traces and the types of algebraic
structures that can describe them.  Since every multiplicity could
admit a symmetric term in front of each distinct color trace, we
should admit symmetric color weights $d^{abc\ldots m}$.  These can be
understood as dressing vertices with $m$ legs.  So $d^{abc}$ can dress
cubic vertices like $f^{abc}$, $d^{abcd}$ dress four-point vertices,
and so on. The combination of various contractions of $f^{abc}$ and
permutation invariant $d$ weights give rise to various algebraic
structures which could have color-dual kinematic weights.  These
structures are rather rich, admitting rules that allow one to generate
a given algebraic structure through functional composition.  When an
algebraic structure depends on scalar kinematics, this can admit a
ladder where composition allows one to climb to higher dimension
effective operators with a small number of primary building blocks
without having to resort to an Ansatz.  At four and five points this
has been shown to close, up to permutation
invariants~\cite{Carrasco:2019yyn, Carrasco:2021ptp}.  Such
compositional approaches have also been generalized to double-trace
representations~\cite{Low:2019wuv, Low:2020ubn}.  Inverting the
relationship between ordered amplitudes and these non-adjoint
kinematic graph weights will induce distinct gauge-invariant
double-copy relationships from the typical KLT formulation.  When both
copies can be organized into into adjoint-type ordered-amplitudes satisfying
KK and BCJ amplitude relations these differences can be pulled into
higher-derivative corrections to a KLT-type
mapping~\cite{Carrasco:2021ptp}.  A general ansatz-based analysis of
higher-derivative generalized KLT mappings has been carried out in
Ref.~\cite{Chi:2021mio} and its relationship to the compositional approach
has been explored in Ref.~\cite{Bonnefoy:2021qgu}.

Finally we point out the surprisingly generality of these ideas.  Moving beyond the types of color structures typically found at tree- and loop-level, one can consider exotic three-dimensional color-dual Chern-Simons type theories~\cite{Ben-Shahar:2021zww}.  The earliest example of such a color-dual theory is Bagger–Lambert–Gustavsson where Ref.~\cite{Bargheer:2012gv} pointed out that, despite the color weights satisfying a three-algebra,  color-dual gauge-theory numerators could be found.  
Fascinatingly, the amplitudes of this theory double copy to those of three-dimensional maximal supergravity theory, which can also be realized as  the adjoint double-copy of the amplitudes of  dimensionally-reduced maximally supersymmetric Yang-Mills theory---a point explored and clarified in Refs.~\cite{Huang:2012vt, Huang:2012wr}. 
Recently, topologically massive amplitudes have also been shown to be color-dual~\cite{Moynihan:2020ejh,Gonzalez:2021bes,Hang:2021oso}, evading  consistency issues that can arise with  massive gauge theories~\cite{Johnson:2020pny} that do not arise from consistent dimensional reduction of massless gauge theories~\cite{Chiodaroli:2015rdg}.

\section{A web of double-copy-constructible theories}
\label{ZoologySection1}

\def\no{\nonumber}
\def\ha{\hat a}
\def\hb{\hat b}
\def\hc{\hat c}
\def\hd{\hat d}
\def\he{\hat e}
\def\hA{\hat A}
\def\hB{\hat B}
\def\hC{\hat C}
\def\hD{\hat D}
\def\hE{\hat E}

\def\tR{t_{\cal R}}

\newcommand{\co}{\ , \ \ \ \ \ \ }
\newcommand{\dd}{\mathrm{d}}
\newcommand{\te}{\textrm}
\newcommand{\al}{\alpha}
\newcommand{\la}{\lambda}
\newcommand{\vph}{\varphi}
\newcommand{\ap}{{\alpha'}}

\newcommand{\haa}{{\hat \alpha}}
\newcommand{\hbb}{{\hat \beta}}
\newcommand{\hgg}{{\hat \gamma}}
\newcommand{\hdd}{{\hat \delta}}
\newcommand{\hee}{{\hat \epsilon}}

\def\cN{{\cal N}}
\def\wone{0.3\textwidth}
\def\wtwo{0.50\textwidth}
\def\wfour{0.14\textwidth}
\def\wthree{0.6\textwidth}
\def\wfive{0.4\textwidth}

\setlength{\LTcapwidth}{\textwidth}

\begin{figure}[t]
\begin{center} 
  \includegraphics[width=1.01\textwidth]{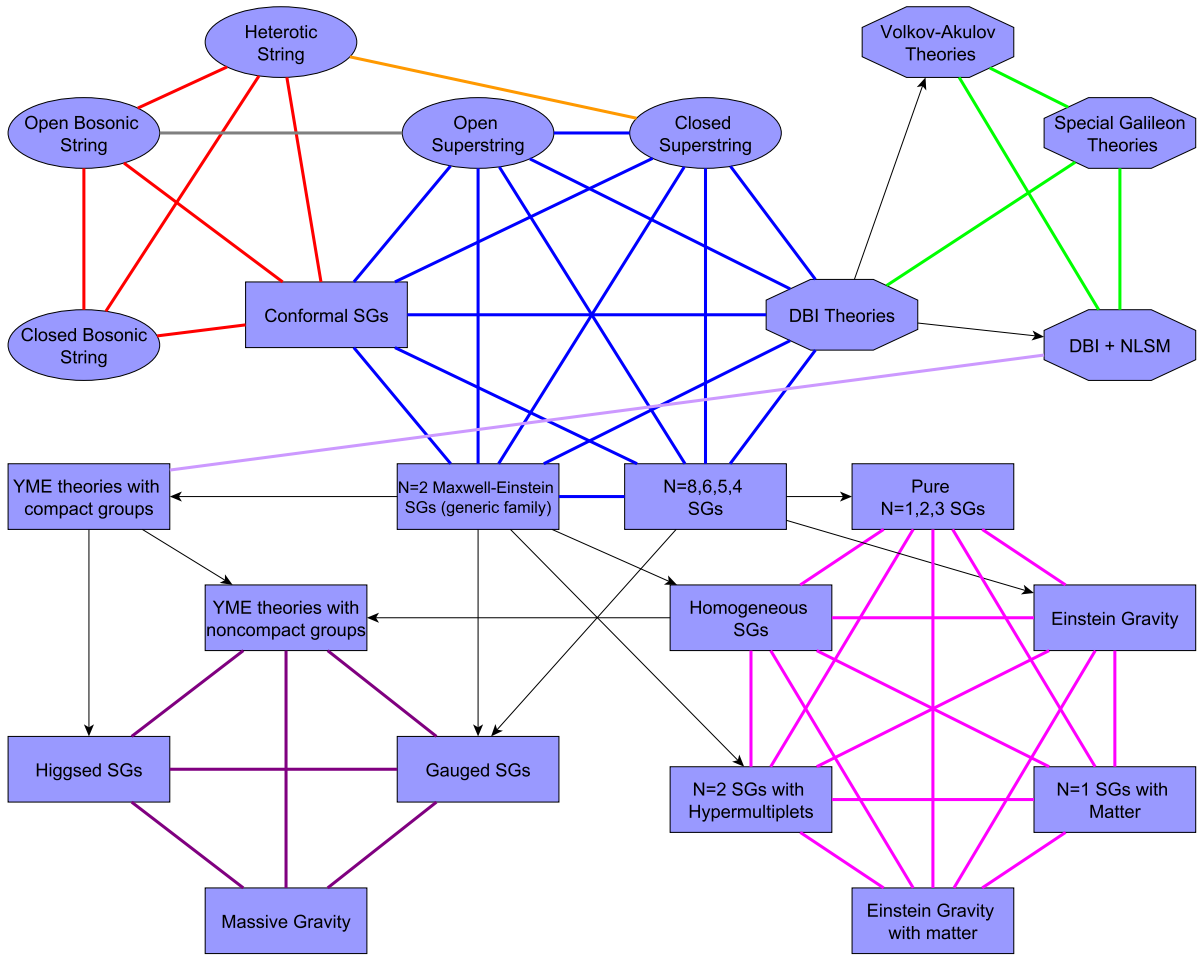} \caption{Web of double-copy-constructible theories. 
 Undirected links with different colors are drawn between theories that have a
 common gauge-theory factor. For example, blue: pure SYM theory, red: $(DF)^2$ theory, green: NLSM, pink: 
 (S)YM theory with massless matter, violet: spontaneously broken (S)YM theory.
 Directed links point toward double-copy constructions that are obtained by modifying both gauge-theory
 factors. Examples include: adding matter representations, assigning VEVs or truncating/projecting out some states. \label{FigWeb}}
\end{center}
\end{figure}

Since their original formulation, color-kinematics duality and the double-copy construction have been applied to a diverse array of theories. 
First, they played a fundamental role in enhancing our understanding of maximal supergravity, particularly in relation to its UV behavior. 
From the beginning, it has also been clear that the double copy can be applied to theories
 that can be interpreted as consistent truncations of maximal supergravity (in some cases, with some subtleties related to the 
 removal of undesired states, for which a variety of methods are now available~\cite{Johansson2014zca, Luna:2016due, Carrasco:2020ywq, Kosmopoulos:2020pcd}).
Many additional examples of double-copy-constructible theories have  emerged, including non-gravitational theories, such as the Dirac-Born-Infeld (DBI) theory, 
and theories which presents structures that are far more involved than maximal supergravity, such as gauged supergravities.
The double copy is now understood as a property of very large  classes of theories, and possibly a generic feature of gravitational interactions. Seemingly-unrelated
theories are now understood to share common building-blocks at the level of the underlying gauge theories entering their double-copy construction. 
We note that some instances of double copy connect string and superstring theories, giving a family of ``stringy''  constructions.   
Similar programs, aiming at connecting different theories in a unified framework, have also  been formulated  in the contexts of the scattering-equations formalism~\cite{Cachazo2014xea}, amplitude transmutation~\cite{CheungUnifyingRelations}, and soft limits~\cite{Cheung:2016drk}.
While we do not have the space to provide a comprehensive summary of all known instances of the double copy, here we aim at giving a broad overview of this web of theories, schematically portrayed in \fig{FigWeb}, as well as an illustration of how the new examples of double copy are connected to the original construction for maximal supergravity.  For further details, we refer the reader to the more detailed review~\cite{BCCJRReview} and to the original literature.

\subsection{Ungauged supergravities}

In order to provide an overview of the available double-copy constructions, we first need to understand how to chart the space of possible gravitational theories. In the presence of supersymmetry, this is a problem that has long been studied by the supergravity community~\cite{Supergravity}.  Supergravity theories can be divided into ungauged theories, Yang-Mills-Einstein (YME) theories, and gauged supergravities.  
The former are theories in which no field is charged under any gauge group. $\cN=8$ (ungauged) supergravity~\cite{Cremmer:1979up} belongs to this group (although several gauged versions are available), together with half-maximal supergravity~\cite{Cremmer:1977tt,Das:1977uy}. These are among the simplest examples of double-copy-constructible theories.   
While ungauged supergravities with $\cN >4$ and two-derivative actions are unique, $\cN=3,4$ two-derivative supergravities are fully specified by a single parameter---the number of 
matter vector multiplets. If we further reduce the number of supercharges, we are in a situation in which not only different kinds of matter multiplets become possible, but additional information about their interactions is needed to  fully specify the theory. This freedom is reflected by the fact that, while extended $\cN>2$ supersymmetry permits only a discrete set of symmetric scalar manifolds, $\cN\leq 2$ supersymmetry is not as constraining.  In four dimensions, supergravities with vector multiplets possess special-K\"{a}hler scalar manifolds, 
while the geometry is  quaternionic-K\"{a}hler in the case of supergravities with hypermultiplets~\cite{Supergravity}. From the double-copy perspective, a particularly important class of 
theories is given by $\cN=2$  Maxwell-Einstein theories that can be uplifted to five dimensions. These theories are fully specified by their vector couplings in five dimensions, i.e. their five-dimensional action includes a term of the form
\begin{equation}
\frac {1}{ 6 \sqrt{6}} C_{IJK} \int F^I \wedge F^J \wedge A^K  \,,
\end{equation}
where the indices $I,J,K$ run over the  vector fields of the theory and $C_{IJK}$ is a constant symmetric tensor. A fundamental result in supergravity states that the supergravity Lagrangian at the two-derivative level can be fully determined once the $C_{IJK}$ tensors are given \cite{Gunaydin:1983bi,Gunaydin:1984ak}. In other words, this class of theories is fully specified by three-point interactions, and hence
constitutes a very convenient arena for applying amplitude methods.

Double-copy constructions with $\cN>4$  are unique; the gauge theory factors are two super-Yang-Mills (SYM) theories with different amounts of supersymmetry \cite{BCJ,BCJLoop,N46Sugra,GravityFour,N5GravFourLoop},
\be 
\cN = (N_1+N_2)  \text{ supergravity}  : \quad  (\cN = N_1 \text{ SYM}) \otimes (\cN = N_2 \text{ SYM}) \, .
\no 
\ee
When we consider $\cN=4$  supergravity, the simplest double copy construction has one free parameter: the number of adjoint scalars in the non-supersymmetric gauge theory \cite{BCJ,BCJLoop,N46Sugra,GravityFour,N5GravFourLoop},
\be 
\cN = 4  \text{ supergravity}  : \quad  (\cN = 4 \text{ SYM}) \otimes ( \text{YM}+ n_s \text{ scalars}) \, .  
\no 
\ee
In turn, this becomes the number of vector multiplets in the outcome of the double copy. 
Color-kinematics duality demands that the couplings between the extra scalars be such that the theory can be regarded as the dimensional reduction of a higher-dimensional pure-YM theory.
Further reducing supersymmetry, the simplest double copy for $\cN=2$ supergravity is of the form \cite{Chiodaroli2014xia}
\begin{equation}
\cN = 2  \text{ supergravity (generic family)} : \quad  (\cN = 2 \text{ SYM}) \otimes ( \text{YM}+ n_s \text{ scalars}) \,.
\no 
\end{equation}
This is the double-copy construction for an infinite family of $\cN=2$ Maxwell-Einstein theories that admit a five dimensional uplift and is known in the literature as the generic family or generic Jordan family. However, this is only one possibility and additional variants of the construction have been formulated. A very important generalization is given by adding matter (half) hypermultiplets to the supersymmetric theory, and matter fermions to the non-supersymmetric theory \cite{Chiodaroli2015wal},
\begin{align}
\left(
\begin{array}{c}
\cN  =   2  \text{ homogeneous} \\
\text{ supergravity }
\end{array} \right)
 : \quad  
\left(
\begin{array}{c}
 \cN = 2 \text{ SYM}  \\
 \null + \frac{1}{2} \text{ hyper}_R
\end{array} \right) 
 \otimes
\left(
\begin{array}{c}
\text{YM}+ n_s \text{scalars}  \\ \null + n_f \text{ fermions}_R
\end{array} \right)  \,. \label{DChom}
\end{align}

\def\spacebefore{\vskip 0.2cm} 
\def\spaceafter{\vskip 0.2cm}
\def\indentspace{\hskip 0.12cm}

\afterpage{
\begin{longtable}{@{}c@{}|c|c}
	{\bf Gravity}	& {\bf Gauge theories} & {\bf Refs.}  \\
	\hline
	\hline
	\parbox{\wone}{ \footnotesize \spacebefore
		$\cN>4$  supergravity  \medskip } &
	\parbox{\wtwo}{ \footnotesize \spacebefore
		$\bullet$ $\cN=4$ SYM theory \\
		$\bullet$ SYM theory ($\cN=1,2,4$)  \spaceafter
	} & \parbox{\wfour}{\footnotesize \spacebefore
\cite{BCJ,BCJLoop,N46Sugra,GravityFour,N5GravFourLoop}} \\
	\hline
	\parbox{\wone}{ \footnotesize \spacebefore
		$\cN=4$  supergravity \\ with  vector  multiplets   \spaceafter } &
	\parbox{\wtwo}{ \footnotesize \spacebefore
		$\bullet$ $\cN=4$ SYM theory \\
		$\bullet$ YM-scalar theory from dimensional  reduction  \spaceafter
	} & \parbox{\wfour}{\footnotesize \spacebefore
	\cite{BCJ,BCJLoop,N46Sugra,N4GravFourLoop}} \\
	\hline
	\parbox{\wone}{ \footnotesize \spacebefore
		$\cN=4$ supergravity \\ with  vector  multiplets   \spaceafter } &
	\parbox{\wtwo}{ \footnotesize \spacebefore
		$\bullet$ $\cN=2$ SYM theory with  hypermultiplets \\
		$\bullet$ $\cN=2$ SYM theory with  hypermultiplets  \spaceafter
	} & \parbox{\wfour}{\footnotesize \spacebefore
		\cite{BCJ,BCJLoop,N46Sugra,N4GravFourLoop}} \\
	\hline
	\parbox{\wone}{ \footnotesize \spacebefore
		pure $\cN<4$ supergravity  \medskip } &
	\parbox{\wtwo}{ \footnotesize \spacebefore
		$\bullet$  (S)YM theory with matter in fundamental rep. \\
		$\bullet$ (S)YM theory with ghosts in fundamental  rep.   \spaceafter 
	} & \footnotesize \cite{Johansson2014zca} \\
	\hline
	\parbox{\wone}{ \footnotesize \spacebefore
	Einstein gravity \medskip } &
		\parbox{\wtwo}{ \footnotesize \spacebefore
			$\bullet$  YM theory with matter in fundamental  rep. \\
			$\bullet$  YM theory with ghosts in fundamental rep.  \spaceafter 
		} & \footnotesize \cite{Johansson2014zca} \\
		\hline
	\parbox{\wone}{ \footnotesize \spacebefore
		$\cN=2$  Maxwell-Einstein \\ supergravities  (generic family)  \spaceafter } &
	\parbox{\wtwo}{ \footnotesize \spacebefore
		$\bullet$ $\cN=2$ SYM theory \\
		$\bullet$ YM-scalar theory from dimensional reduction  \spaceafter
	} & \footnotesize \cite{Chiodaroli2014xia}  \\
	\hline
	\parbox{\wone}{ \footnotesize \spacebefore
		$\cN=2$  Maxwell-Einstein \\ supergravities (magical/ \\ homogeneous theories)  \spaceafter } &
	\parbox{\wtwo}{ \footnotesize \spacebefore
		$\bullet$ $\cN=2$ SYM theory with half   hypermultiplet in \\ $\vphantom{x}$ \indentspace pseudoreal representation\\
		$\bullet$ YM-scalar theory from dimensional  reduction 
		 \\ $\vphantom{x}$ \indentspace with  matter fermions in pseudo-real representation  \spaceafter 
	} & \footnotesize \cite{Chiodaroli2015wal,Ben-Shahar:2018uie}  \\
	\hline		
	\parbox{\wone}{ \footnotesize \spacebefore
		$\cN=2$  supergravities  \\ with  hypermultiplets  \spaceafter } &
	\parbox{\wtwo}{ \footnotesize \spacebefore
		$\bullet$ $\cN=2$ SYM theory with half hypermultiplet \\
		$\bullet$ YM-scalar theory from dimensional  reduction   
		\\ $\vphantom{x}$ \indentspace with extra matter scalars  \spaceafter 
	} & \footnotesize \cite{Chiodaroli2015wal,Anastasiou2017nsz} \\
	\hline		
	\parbox{\wone}{\footnotesize \spacebefore
		$\cN=2$  supergravities  \\ with   vector/ \\ hypermultiplets  \spaceafter } &
	\parbox{\wtwo}{ \footnotesize  \spacebefore
		$\bullet$ $\cN=1$ SYM theory with chiral multiplets \\
		$\bullet$ $\cN=1$ SYM theory with chiral  multiplets  \spaceafter 
	} & 
    \footnotesize 
	\cite{OneLoopSusy,Damgaard2012fb,Anastasiou2015vba}
     \\
	\hline	
	\parbox{\wone}{ \footnotesize \spacebefore
		$\cN=1$   supergravities with \\ vector multiplets \\  (truncations  of generic  family)  \spaceafter } &
	\parbox{\wtwo}{ \footnotesize \spacebefore
		$\bullet$ $\cN=1$ SYM theory \\
		$\bullet$ YM-scalar theory from dimensional red. 
	} & \footnotesize \cite{Chiodaroli2014xia}  \\ 
	\hline
	\parbox{\wone}{ \footnotesize \spacebefore
		$\cN=1$  supergravities  with \\  vector multiplets  
		(truncations \\ of homogeneous theories) \spaceafter } &
	\parbox{\wtwo}{ \footnotesize \spacebefore
		$\bullet$ $\cN=1$ SYM theory with chiral multiplets \\ $\vphantom{x}$ \indentspace 
		in fundamental representation \\
		$\bullet$ YM-scalar theory with fermions \\ $\vphantom{x}$ \indentspace 
		in fundamental representation  \spaceafter 
	} & 
    \parbox{\wfour}{\footnotesize \spacebefore
    	\cite{OneLoopSusy,Damgaard2012fb,Anastasiou2015vba,Johansson2014zca}  \spaceafter }
    \\
	\hline		
	\parbox{\wone}{ \footnotesize \spacebefore
		$\cN=1$  supergravities \\ with chiral multiplets  \spaceafter } &
	\parbox{\wtwo}{ \footnotesize \spacebefore
		$\bullet$ $\cN=1$ SYM theory with chiral multiplets  \\ $\vphantom{x}$ \indentspace in fundamental representation \\
		$\bullet$ YM-scalar with extra  scalars in fundamental rep.  \spaceafter
	} & 
    \parbox{\wfour}{\footnotesize \spacebefore
    	\cite{OneLoopSusy,Damgaard2012fb,Anastasiou2015vba,Johansson2014zca} \spaceafter }
 \\
	\hline		
	\parbox{\wone}{ \footnotesize \spacebefore
		Einstein gravity \\ with massless matter  \spaceafter } &
	\parbox{\wtwo}{ \footnotesize \spacebefore
		$\bullet$ YM theory with matter \\
		$\bullet$ YM theory with matter  \spaceafter
	} & \footnotesize \cite{BCJ,Johansson2014zca}   \\
	\hline
	\parbox{\wone}{ \footnotesize \spacebefore
		Einstein gravity \\ with massive scalars  \spaceafter } &
	\parbox{\wtwo}{ \footnotesize \spacebefore
		$\bullet$ massive scalar QCD \\
		$\bullet$ massive scalar QCD  \spaceafter
	} & \footnotesize \cite{Plefka:2019wyg,Carrasco:2020ywq}   \\
	\hline
	\parbox{\wone}{ \footnotesize \spacebefore
		Heavy-mass \\ effective theory  \spaceafter } &
	\parbox{\wtwo}{ \footnotesize \spacebefore
		$\bullet$ heavy-quark effective theory \\
		$\bullet$ heavy-quark effective theory  \spaceafter
	} & \footnotesize \cite{Haddad:2020tvs,Brandhuber:2021kpo}   \\
	\hline		
	\parbox{\wone}{\footnotesize \spacebefore
		Einstein gravity with \\
		higher-derivative  corrections  \spaceafter } &
	\parbox{\wtwo}{ \footnotesize \spacebefore
		$\bullet$ YM theory with higher-derivative corrections \\
		$\bullet$ YM theory with higher-derivative corrections  \spaceafter 
	} & \footnotesize \cite{Broedel2012rc,Carrasco:2019yyn,Chi:2021mio} \\
	\hline		
	\parbox{\wone}{\footnotesize \spacebefore
		Massive gravity/ \\ Kaluza-Klein gravity  \spaceafter } &
	\parbox{\wtwo}{ \footnotesize \spacebefore
		$\bullet$ spontaneously-broken YM theory \\
		$\bullet$ spontaneously-broken YM theory  \spaceafter 
	} & \footnotesize \cite{Chiodaroli:2015rdg,Momeni:2020vvr,Momeni:2020hmc,Johnson:2020pny} \\
	\hline		
	\parbox{\wone}{ \footnotesize \spacebefore
		$\cN\leq4$ Conformal \\ (super)gravity  \spaceafter } &
	\parbox{\wtwo}{ \footnotesize \spacebefore
		$\bullet$ $DF^2$ theory  \\
		$\bullet$ (S)YM theory   \spaceafter 
	} & \footnotesize \cite{Johansson2017srf,JohanssonConformal,Menezes:2021dyp} \\
		\hline
		\parbox{\wone}{ \footnotesize \spacebefore
			3D maximal \\ supergravity  \spaceafter} &
		\parbox{\wtwo}{ \footnotesize \spacebefore
			$\bullet$ BLG theory  \\
			$\bullet$ BLG theory   \spaceafter 
		} & 
	 \footnotesize
	 	\cite{Huang:2012wr,Bargheer:2012gv,AllicABJM}\\
		\hline 
\multicolumn{3}{c}{}\\
	\caption{ 	\small  Non-exhaustive list  of ungauged double-copy-constructible gravitational theories presented in the literature with references. Theories are specified in four dimensions (with the exception of the last entry). \label{table-zoology-ungauged}}
\end{longtable}}

For technical reasons, the matter representation is taken to be pseudo-real, which makes it possible to introduce a single half-hypermultiplet in the supersymmetric theory. Since we are in the presence of more than one type of gauge-group representation, 
we need to generalize color-kinematics duality beyond the purely-adjoint case, as we have already seen in Sec. \ref{nonadjck}.
This is done according to the following rule: 

\smallskip

\begin{center}
	\parbox{0.9\textwidth}{ \it
		Numerator factors in a CK-duality-satisfying presentation of a gauge-theory amplitude obey the same algebraic relations as the color factors. This includes those relations which stem from Jacobi identities or commutation relations of gauge group generators, as well as additional relations that are required by gauge invariance. }
\end{center}

\smallskip

\noindent
Additionally, we need to decide how different representations are combined by the double copy. To this end, we can use a simple and elegant working rule:

\smallskip

\begin{center}
	\parbox{0.9\textwidth}{ \it
		 Each state in the double-copy (gravitational) theory corresponds to a gauge-invariant bilinear of gauge-theory states. }
\end{center}

\smallskip

\noindent
For this to be possible, we identify the  gauge groups of the two theories entering the construction.  Concretely, this rule implies that a supergravity field is obtained by combining two adjoint or two matter gauge-theory fields, but no supergravity field can originate from the double copy of one adjoint and one matter field, since this combination cannot form a gauge singlet. Because of this rule, the double copy (\ref{DChom}) does not contain any additional gravitino multiplets, and the contribution of the extra matter fields simply yields additional vector multiplets. Furthermore, the number of matter fermions $n_f$ is constrained by the requirement that the gauge theory should be seen as a higher-dimensional YM theory with fermions. This requirement is a consequence of color-kinematics duality, and the reader is referred to Ref.~\cite{Chiodaroli2015wal} for the full analysis. Taking this constraint into account, we have a two-parameter family of double copies which perfectly matches the classification of $\cN=2$ Maxwell-Einstein supergravities with homogeneous scalar manifolds that has been obtained in the supergravity literature~\cite{deWit1991nm}.  

There are many more examples of double-copy constructions giving ungauged supergravities. A particularly important one is the construction for Einstein gravity. Simple counting of states shows that the double copy of two pure YM theories yields additional states beyond those of the graviton (in four dimensions, an additional complex scalar corresponding to a dilaton and an axion). An interesting way to remove the unwanted states is to introduce matter fermions in one of the two YM theories and matter ghost fields in the other~\cite{Johansson2014zca}. These fields only double copy with each other in accordance to the rule given before. Only amplitudes with external gravitons are considered so that matter fields and ghosts appear only in loops.  Ref. \cite{Johansson2014zca} shows that the loop contributions coming from ghost fields are precisely what is needed to cancel 
the contribution of the  unwanted axion-dilaton degrees of freedom, resulting in a double-copy construction for pure Einstein gravity. One can also use physical-state projectors to remove the unwanted states, as done in, for example, Ref.~\cite{3PMLong}.

Given its role in the constructions outlined in this section, the
reader may wonder whether SYM theory is the only purely-adjoint theory
that obeys color-kinematics duality.  It turns out that there is
another theory with this property, which also appears in several
double-copy constructions. This is the so-called $(DF)^2$ theory,
which, in its simplest incarnation, is a higher-derivative version of
the YM theory with a mass parameter $m$. It has Lagrangian
\begin{align}
	{\cal L}_{(DF)^2+{\rm YM}}&= \frac{1}{2}(D_{\mu} F^{a\, \mu \nu})^2- \frac{1}{4} m^2 (F^a_{\mu \nu})^2 \,.
\end{align}
This minimal version of the $(DF)^2$  theory enters the double-copy construction for a mass deformation of conformal supergravity,
\be
\big( \text{mass-deformed minimal CSG} \big)=\big({\rm SYM}\big) \otimes \big(\textrm{minimal  $(DF)^2$ + YM}\big)\, .
\label{CG_DC_massdef}
\ee
The above construction gives amplitudes in a mass-deformed minimal $\cN=4$ theory that interpolates between (Weyl)${}^2$ and a Ricci-scalar term. 
Supersymmetry can be reduced by modifying the first gauge-theory factor. Additionally, this $(DF)^2$ theory  has also a non-minimal version,  
containing an $F^3$ term  together with further ghost scalars transforming in a specific matter representation.
In Table \ref{table-zoology-ungauged}, we summarize double-copy constructions giving ungauged gravitational theories, and include 
references to the original literature.

\subsection{Yang-Mills-Einstein and gauged supergravities with Minkowski vacua}

YME theories and gauged supergravities are supergravity theories that contain gauge interactions under which some of the  fields are charged. The 
YME theories are obtained by promoting a non-abelian subgroup of the global isometry group of a Maxwell-Einstein supergravity to a local symmetry (without touching the $R$ symmetry and without introducing additional fields). In contrast, the defining property of gauged supergravities is that part of the $R$ symmetry is promoted to gauge symmetry. These theories are considerably more involved than their YME relatives, exhibiting, among other things, non-trivial potentials, spontaneously-broken supersymmetry and massive gravitini. 
The reader interested in the relevant supergravity literature may consult Refs.~\cite{Supergravity,Samtleben2008pe}.
Amplitudes in YME theories have been intensely investigated with a variety of methods: scattering equations~\cite{Cachazo2014nsa,Cachazo2014xea,Nandan2016pya}, collinear limits~\cite{Stieberger:2015qja}, on-shell recursion~\cite{TengFengBCJNumerators,Du2017gnh},  string theory~\cite{Stieberger2016lng,SchlottererEYMHeterotic} and ambitwistor strings \cite{Casali2015vta}.
From the point of view of the double-copy construction~\cite{Chiodaroli2014xia}, 
non-abelian gauge interactions in the double-copy theory are generated by introducing a trilinear coupling among 
the adjoint scalar fields in the non-supersymmetric gauge-theory factor.
These coupling are written as 
\begin{equation}
\delta {\cal L} =
\frac{\lambda} {6!} F^{IJK}  {\rm Tr} [\phi^I, \phi^J ] \phi^K \ ,
\end{equation}
where $F^{IJK}$ is an antisymmetric tensor with indices running over the number of scalars in the theory. The effect of these couplings is to introduce non-zero supergravity 
amplitudes between three vectors which are proportional to the $F^{IJK}$ tensors. In turn, imposing color/kinematics duality on amplitudes between four scalars is equivalent to requiring that these tensors  obey Jacobi relations, and hence can be thought of as the structure constants of  the supergravity gauge group. This is an example of a global symmetry in a gauge-theory factor being promoted to a local symmetry by the double copy, analogous to the relation between global and local supersymmetry. 
The net result is a double-copy of the form \cite{Chiodaroli2014xia}
\begin{equation}
\big( \text{YME supergravity} \big) : \quad \big( \text{SYM theory} \big) \otimes \big( \text{YM} + \phi^3 \text{ theory}  \big)  ,
\end{equation}
where, in case of $\cN=2$, the $\lambda \rightarrow 0$ limit will yield a theory belonging to the generic family. YME theories with spontaneously-broken gauge groups can also be constructed by taking the SYM gauge theory on its Coulomb branch and introducing extra massive scalars in the non-supersymmetric theory while making sure that color-kinematics duality is preserved~\cite{Chiodaroli:2015rdg}.  
  
\begin{table}[t]
	\begin{tabular}{@{}c@{}|c|c}
		{\bf Gravity}	& {\bf Gauge theories} & {\bf Refs.}  \\
		\hline
		\hline
		\parbox{\wone}{ \footnotesize \spacebefore
			Unbroken $\cN\leq4$  Yang-Mills-\\Einstein  supergravities \spaceafter } &
		\parbox{\wtwo}{ \footnotesize \spacebefore
			$\bullet$ SYM theory \\
			$\bullet$ YM-scalar theory with trilinear  scalar couplings   \spaceafter
		} & \parbox{\wfour}{\footnotesize \spacebefore \centering
			\cite{Bern1999bx,Chiodaroli2014xia,Cachazo2014nsa} \\  \cite{Cachazo2014xea,Nandan2016pya,TengFengBCJNumerators} \\  \cite{Du2017gnh,Stieberger2016lng,Casali2015vta,CheungUnifyingRelations,SchlottererEYMHeterotic} } \\
		\hline		
		\parbox{\wone}{ \footnotesize \spacebefore
			Higgsed $\cN\leq4$  Yang-Mills-\\ Einstein supergravities  \spaceafter } &
		\parbox{\wtwo}{ \footnotesize \spacebefore
			$\bullet$ SYM theory on the Coulomb branch\\
			$\bullet$ YM-scalar theory with trilinear scalar couplings \\  $\vphantom{x}$ \indentspace and extra  massive scalars  \spaceafter
		} & \footnotesize \cite{Chiodaroli:2015rdg}  \\
		\hline
		\parbox{\wone}{ \footnotesize \spacebefore
			$\cN=2$  Yang-Mills-\\ Einstein  supergravities \\  
			(non-compact  gauge groups) \spaceafter} &
		\parbox{\wtwo}{ \footnotesize \spacebefore
			$\bullet$ $\cN=2$ SYM theory on the Coulomb branch \\  $\vphantom{x}$ \indentspace with massive hypers\\
			$\bullet$ YM-scalar theory with trilinear scalar couplings 
			\\  $\vphantom{x}$ \indentspace and massive  fermions
			\spaceafter}
		& \footnotesize \cite{Chiodaroli:2022next}  \\
		\hline
		\parbox{\wone}{ \footnotesize \spacebefore
			$U(1)_R$ gauged  supergravities \\(with Minkowski vacua) \spaceafter} &
		\parbox{\wtwo}{ \footnotesize \spacebefore
			$\bullet$ SYM theory on Coulomb branch\\
			$\bullet$ YM theory with SUSY broken by fermion masses \spaceafter
		} & \footnotesize \cite{Chiodaroli2017ehv}  \\
		\hline

		\parbox{\wone}{ \footnotesize \spacebefore
			Non-abelian gauged \\ supergravities \\ (with Minkowski vacua) \spaceafter} &
		\parbox{\wtwo}{ \footnotesize \spacebefore
			$\bullet$ SYM theory on the Coulomb branch\\
			$\bullet$ YM-scalar theory with trilinear scalar couplings \\ $\vphantom{x}$ \indentspace and massive  fermions
			\spaceafter}
		& \footnotesize \cite{Chiodaroli:2018dbu}  \\
		\hline 
	\end{tabular}
	\caption{Gauged/YME gravities and supergravities for which a double-copy construction 
		is presently known.  \label{table-zoology-gauged}}  	
\end{table}

Gauged supergravities, even those admitting Minkowski vacua, are considerably more involved. Their double-copy construction
can be thought of as a generalization of the construction for YME theories in which a spontaneously-broken YM theory is combined with a theory in which supersymmetry is broken by explicit fermionic masses.\footnote{The double-copy description of gauged supergravities in non-Minkowski vacua is an open problem.}  As in the construction for YME theories, the appearance of trilinear scalar couplings results in non-abelian interactions in the supergravity theory, but now the $F$-tensors are also related to the fermionic masses by color-kinematics duality.
The study of gauged supergravities in the double-copy language is still in its infancy, and the reader should consult  Refs. \cite{Chiodaroli2017ehv} and  \cite{Chiodaroli:2018dbu} for additional details.
The presently-known double-copy constructions for Yang-Mills-Einstein theories and gauged supergravities are listed in Table \ref{table-zoology-gauged}.   
Various theories without a graviton, most prominently variants of the DBI theory, have also been shown to admit such construction (see Table \ref{table-zoology-nongrav} for an overview).

\begin{table}[t]
	\begin{center}
		\begin{tabular}{@{}c@{}|c|c}
			{\bf Double copy}	& {\bf Starting theories} & {\bf Refs.}  \\
			\hline
			\hline
			\parbox{\wone}{ \footnotesize \spacebefore
				$\cN\leq4$ Dirac-Born-Infeld \\ theory} &
			\parbox{\wtwo}{ \footnotesize \spacebefore
				$\bullet$ NLSM  \\
				$\bullet$ (S)YM theory  \\ } 
			&\footnotesize 	\parbox{\wfour}{ \centering  \cite{Cachazo2014xea,CheungUnifyingRelations,Chen2013fya,Du2016tbc,Chen2014dfa,Chen2016zwe,Cheung:2017yef}}
			\\
			\hline		
			\parbox{\wone}{ \footnotesize \spacebefore
				Volkov-Akulov   theory} &
			\parbox{\wtwo}{ \footnotesize \spacebefore
				$\bullet$ NLSM  \\
				$\bullet$ SYM theory (only fermions as external states)  \\ } 
			&   		\parbox{\wfour}{\centering \footnotesize \cite{Bergshoeff:1996tu,Bergshoeff:1997kr,Kallosh:1997aw,Bergshoeff:2013pia,Cachazo2014xea}\\  \cite{Cachazo2016njl, He2016mzd, Elvang:2018dco} }
			\\
			\hline		
			\parbox{\wone}{ \footnotesize \spacebefore
				Special Galileon theory \spaceafter} &
			\parbox{\wtwo}{ \footnotesize \spacebefore
				$\bullet$ NLSM  \\
				$\bullet$ NLSM  \\ 
			} & 	\parbox{\wfour}{ \centering \footnotesize \cite{Cachazo2014xea,CheungUnifyingRelations,Cheung:2016prv} \\  \cite{Cachazo2016njl,Cheung:2017yef}  }
			\\
			\hline
			\parbox{\wone}{ \footnotesize \spacebefore
				$\cN \leq 4$	DBI  + (S)YM   \\  theory } &
			\parbox{\wtwo}{ \footnotesize \spacebefore
				$\bullet$ NLSM + $\phi^3$  \\
				$\bullet$ (S)YM theory   \\ 
			} &  		\parbox{\wfour}{ \centering	\footnotesize \cite{Chiodaroli2017ngp,Cachazo2014xea,CheungUnifyingRelations,Chen2013fya,Du2016tbc,Chen2014dfa,Chen2016zwe,Cachazo2016njl,Carrasco2016ygv}} 
			\\
			\hline
			\parbox{\wone}{ \footnotesize \spacebefore
				DBI + NLSM  \\   theory } &
			\parbox{\wtwo}{ \footnotesize \spacebefore
				$\bullet$ NLSM   \\
				$\bullet$ YM + $\phi^3$ theory   \\ 
			} &  		\parbox{\wfour}{ \centering	\footnotesize \cite{Chiodaroli2017ngp,Cachazo2014xea,CheungUnifyingRelations,Chen2013fya,Du2016tbc,Chen2014dfa,Chen2016zwe}} 
			\\
			\hline
			\parbox{\wone}{ \footnotesize \spacebefore
				3D $\cN=8$ \\ DBI theory  \spaceafter} &
			\parbox{\wtwo}{ \footnotesize \spacebefore
				$\bullet$ 3D $\cN=4$ Chern-Simons-matter theory  \\
				$\bullet$ 3D $\cN=4$ Chern-Simons-matter theory   \spaceafter 
			} & 
			\footnotesize
			\cite{Ben-Shahar:2021zww}\\
			\hline 							
		\end{tabular}
		\caption{Non-gravitational local field theories constructed as double copies.  
			\label{table-zoology-nongrav}}
	\end{center}
\end{table}

\subsection{Stringy double copies}  

An important family of double-copy constructions applies to string-theory amplitudes. In this case, a fundamental ingredient is given by a set of disk integrals with punctures \cite{MafraBCJAmplString,Broedel2013tta},
\be
Z_\sigma(\rho(1,2,\ldots,n)) = (2 \ap)^{n-3} \! \! \! \! \! \! \! \! \! \! \! \! \! \! \! \! \! \! \! \!  \! \! \! \!\int \limits_{\sigma \, \{  - \infty \leq z_{1}\leq z_{2} \leq \ldots \leq z_{n}\leq \infty \}} \! \! \! \! \! \! \! \! \! \! \! \! \! \! \! \! \! \! \! \! \frac{d z_1 \, \ldots\, d z_{n}}{{\rm vol}({\rm SL}(2,\mathbb{R}))} \ 
\frac{ \prod_{i<j}^{n} |z_{ij}|^{\alpha' s_{ij}}  }{ \rho \, \{ z_{12} z_{23} \cdots z_{n-1,n} z_{n,1} \} } \, .\label{discint}
\ee
We use the short-hand notation $z_{ij}=z_i-z_j$, and we take care of the  ${\rm vol}({\rm SL}(2,\mathbb R)$ factor by fixing three punctures as $z_i,z_j,z_k\rightarrow (0,1,\infty) $ while introducing a Jacobian $|z_{ij}z_{ik} z_{jk}|$. The above integrals depend explicitly on two permutations $\sigma,\rho \in S_n$. They  are known to satisfy~\cite{Broedel2013tta} field-theory  BCJ relations to all multiplicity with respect to the permutation~$\rho$,
\be 
\sum_{j=2}^{n-1}  (p_1\cdot p_{23\ldots j})Z_\sigma (2,3,\ldots,j,1,j+1, \ldots, n) = 0 \,,
\label{ftarel}
\ee
and the so-called string-theory monodromy relations~\cite{Monodromy,Stieberger:2009hq} with respect to $\sigma$,
\be \sum_{j=1}^{n-1}  e^{ 2 i \pi \alpha' p_1\cdot p_{23\ldots j}}Z_{(2,3,\ldots,j,1,j+1, \ldots, n)}(\rho) = 0 \,.
\ee
Having introduced the appropriate building blocks, the open-superstring amplitudes with color-ordered massless external states can be expressed as the double copy of the $Z$ integrals with   Yang-Mills scattering amplitudes~\cite{MafraBCJAmplString,Broedel2013tta},
\begin{equation}
A^{\tree}_{\rm OS}(\sigma(1,2,3,\ldots,n)) ~ =\!\!\!\!\!\!\!\!\!  \sum_{\tau,\rho \in S_{n-3}(2,...,n-2)}\!\!\!\!\!\!\!\!\!  Z_\sigma (1,\tau,n,n{-}1)  S[\tau | \rho]  A_{\rm SYM}(1,\rho,n{-}1,n) \,,
\label{2.2cOLD} 
\end{equation}
where the field-theory KLT kernel $S[\tau|\rho]$ has been introduced
in Eq.  (\ref{momKernel}).  

The $Z$ integrals have been interpreted as
the amplitudes of a scalar theory dubbed Z-theory in Refs.~\cite{Carrasco2016ldy, Mafra2016mcc, Carrasco2016ygv}.  While it is
surprising that the field-theory version of the KLT kernel appears
here, this may be understood from the fact that the decomposition is in terms of
SYM amplitudes that obey field-theory BCJ relations.  It is
remarkable that in the superstring all the $\alpha’$ dependence is
contained in the Z theory.  A closed-string version of the Z-theory
integrals, known to also satisfy field-theory relations to all multiplicity, is given by the following integrals on the punctured Riemann
sphere
\cite{Stieberger2014hba,Schlotterer2018zce,Vanhove:2018elu,Brown2018omk},
\be
{\rm sv} \, Z(\tau  | \sigma) =  \left( \frac{2\ap}{\pi } \right)^{n-3}  \! \! \!  \int  \frac{d^2 z_1  \, \ldots\, d^2 z_{n}}{{\rm vol}({\rm SL}(2,\mathbb{C}))} \ 
\frac{ \prod_{i<j}^{n} |z_{ij}|^{2 \alpha' s_{ij}}  }{ \tau \, \{ \bar{z}_{12} \bar{z}_{23} \cdots \bar{z}_{n-1,n} \bar{z}_{n,1} \}  \sigma \, \{ z_{12} z_{23} \cdots z_{n-1,n} z_{n,1} \} } \,.
\label{2.2bcl}
\ee
The notation ${\rm sv} Z$ refers to the so-called single-valued projection  of multiple zeta values (MZVs) (see Refs.~\cite{Schnetz2013hqa, Brown2013gia} for details), but for us it will be simply part of the name of the building blocks we are introducing. Using these integrals,  closed-superstring amplitudes are schematically given as \cite{Schlotterer:2012ny,Stieberger:2013wea}
\be
(\text{closed superstring})= ({\rm SYM}) \otimes  {\rm sv}\big(\textrm{open superstring} \big) \, .
\label{CSvsOS}
\ee
The known stringy double copies are summarized in Table~\ref{stringOverview}. Note that the $(DF)^2$ theory we have introduced in the beginning of this section appears (in its non-minimal version) in several of the entries. 

Each column in Table~\ref{stringOverview} corresponds to the computation of one type of correlator.  The SYM column is derived for any number of external massless states~\cite{MafraBCJAmplString}. The $(DF)^2+{\rm YM}$ and the $(DF)^2+{\rm YM}+\phi^3$ columns have been explicitly checked against string amplitudes through five points, and all-multiplicity arguments were also given in Ref.~\cite{Azevedo:2018dgo}.  While the discussion here focuses on tree-level amplitudes, some extensions to loop level are available in the literature~\cite{MafraSchlotterOneLoopString,Mafra:2012kh,Mafra2018nla,Mafra2018pll,Mafra2018qqe,Mafra:2019ddf,Mafra:2019xms,DHoker:2020prr,DHoker:2020tcq,Gerken:2020xfv,Stieberger:2021daa,Edison:2021ebi,DHoker:2021kks}.  
See also Refs.~\cite{He:2018pol,He:2019drm} for a construction of string amplitudes in terms of field-theory amplitudes using  the scattering-equations formalism.

\begin{table}[tb]\small
	\be \! \! \!
	{
		\begin{array}{c|c|c|c}
			{\rm string}\otimes {\rm QFT}  &\te{SYM} &(DF)^2 \,{+}\, \te{YM}&\ (DF)^2\, {+}\, \te{YM}\,{+}\,\phi^3 \\\hline \hline
			\te{Z-theory} \ & \ \te{open superstring} \ \, & \, \te{open bosonic string} \, &{ \begin{array}{c} 
					\te{compactified open} \\
					\te{bosonic string}
			\end{array}}  \\
			\te{sv}(\te{open superstring})  \ & \ \te{closed superstring}  \ \, &\ \te{heterotic (gravity)} \ \, &\, \te{heterotic\,(gauge/gravity)}  \\
			\te{sv}(\te{open\,bosonic\,string})  \, &  \, \te{heterotic\,(gravity)} \, & \, \te{closed\,bosonic\,string} \, &{ \begin{array}{c} 
					\te{compactified closed} \\
					\te{bosonic string}
			\end{array}}
	\end{array}} \nonumber
	\ee
	\caption{Double-copy constructions of tree-level string amplitudes with external massless states~\cite{Azevedo2018dgo}. The single-valued projection sv($\bullet$) converts the disk integrals (\ref{discint}) to sphere integrals (\ref{2.2bcl}). }
	\label{stringOverview}
\end{table}  


\section{From amplitudes to gravitational waves through the double copy}
\label{ClassicalDoubleCopySection}

\def\eqref#1{(\ref{#1})}

Previous sections have outlined a new approach to a wide class---perhaps even all---gravitational theories, in which they are 
obtained  from simpler gauge theories. Applied beyond scattering amplitudes, similar procedures 
have been shown to relate certain classes of solutions of Einstein's equations to solutions of Maxwell's\footnote{These solutions 
can be thought of as being embedded in Yang-Mills solutions, by giving a nontrivial profile to the vector corresponding to a single 
generator of the gauge group.} 
equations with sources, a simple example of which is the Schwarzschild solution~\cite{Monteiro2014cda}.
Since this method has been used for nontrivial calculations of
supergravity ultraviolet properties up to five loops (see
e.g.~Refs.~\cite{GravityThree, Bern:2012cd,N4GravFourLoop,N5GravFourLoop,
UVFiveLoops}), it is logical to suspect that it can be useful to also
advance the state of the art in gravitational-wave physics based on
Einstein's general relativity by carrying out calculations that are
difficult through standard methods. 
A good choice is high orders of two-body classical gravitational dynamics, given that it feeds into
the analysis of gravitational-wave signals from the LIGO/Virgo
collaborations~\cite{Abbott:2016blz} and is of interest to LIGO
theorists~\cite{Damour:2017zjx}.

Scattering amplitudes and associated methods enter the picture through
the observation that, up to a point, scattering and bound-state motion
are governed by the same equations of motion and Hamiltonian. Thus one may find the
Hamiltonian from a scattering analysis and use it subsequently for
analyzing bound-state motion.\footnote{This philosophy requires care
and possible modifications at ${\cal O}(G^4)$, where the Hamiltonian
depends on the trajectory through the so-called tail effect~\cite{Thorne:1980ru}, so a
scattering-based Hamiltonian cannot be directly applied to bound-state
problems.} The double copy enters very directly, because of 
its natural use in scattering processes.
This strategy of effectively integrating out gravitons carrying momenta responsible for long-range interactions yields a 
two-body Hamiltonian, and can in principle be extended to the construction of $n$-body Hamiltonians.  
Such Hamiltonians can be interpreted as generating functions of classical observables.

To this end, we model the various classical bodies as point-particles,
with or without spin depending on whether or not the classical bodies are 
spinning. This is a reasonable approximation if they are
sufficiently far apart and may be systematically corrected to account
for finite-size effects~\cite{Goldberger:2004jt}.
We begin by reviewing the kinematics, scale hierarchies, power
counting, and truncation of graph structures that allow us to identify
and remove the quantum contributions at the integrand level. Because
of the macroscopic nature of the scattering bodies, it will turn out
that loop-level amplitudes contain classical physics. The methods
reviewed below lead to simplifications which we will also illustrate
and are important for success at high loop orders.

\subsection{Matter and graviton kinematics and the classical limit \label{MatterGravitonKinematics}}

There are several ways to extract classical physics from quantum field
theory and more specifically from scattering amplitudes. We will use
the correspondence principle---that is that classical physics emerges
from the quantum physics in the limit of large masses and
charges. Chief among them is the angular momentum: to extract
the classical part of a four-point elastic amplitude we must therefore
select a kinematic configuration in which the angular momentum is
large in natural ($\hbar=1$) units \cite{CheungPM, 3PM, 3PMLong}.
It is not difficult to see that this implies the more intuitive
picture that classical physics governs processes in which the minimal
inter-particle separation is much larger than the de~Broglie wavelength,
$\lambda$, of each particle. Indeed,
\be
        J\sim |\bm b \times \bm p|\gg1
        \qquad\Rightarrow\qquad
        	|\bm b| \gg \lambda = \frac{1}{|\bm p |}  \, .
\ee
For a scattering process we may take the impact parameter $|\bm b|$ as a measure of the minimal separation, while for a bound 
state we may take it to be the periastron or the average radius for quasi-circular orbits. 

Since the impact parameter is of order of the inverse momentum transfer in a scattering process, $|\bm b| \sim 1/|\bm q|$, the
classical limit implies the kinematic hierarchy\footnote{This
  hierarchy implies that our results should not be expected to be
  valid for massless particles. Indeed, one can see that the classical and massless limits do not
  commute \cite{3PMLong}.  }
\be
	m_1, m_2 , |\bm p| \sim J\,  |\bm q| \, \gg |\bm q|\,.
\label{eq:classical_limit}
\ee
Classical and quantum contributions to scattering processes enter at
different orders in an expansion in large $J$, or equivalently, in
small $|\bm q|$. For example, since Newton's potential is
classical, it follows that in the limit \eqn{eq:classical_limit} any generating function of
classical observables (e.g. the effective potential, the eikonal, the
radial action, etc.) for scalar bodies has the general form
\be
V = \frac{G}{|\bm q|^2}c_1(\bm p)+ \frac{1}{|\bm q|^3} \sum_{n\ge 2} {(G |\bm q|)^n (\ln \bm q^2 )^{n \text{ mod 2}}  c_n(\bm p)} \, .
\label{potential}
\ee
For spinning bodies this expression is augmented with a
dependence on scalars constructed from an equal numbers of the
 transferred momentum vector $\bm q$ and the rest
frame spin $\bm S/m$~\cite{Bern:2020buy}.
Quantum corrections can be systematically included by keeping terms with suitably subleading $\bm q$ counting. 

We note that a small momentum transfer, as in Eq.~\eqref{eq:classical_limit}, is not in contradiction with the observation 
that motion on a closed orbit required a change in momentum of a particle that is comparable with its initial momentum. Indeed,
such long-term classical processes compound a large number of elementary two-particle interactions mediated by graviton exchange.
Each such interaction transfers a momentum $|\bm q|$ compatible with \eqn{eq:classical_limit} while the complete classical process 
transfers a momentum commensurate with $|\bm p|$. In the case of scattering, this is concretely described 
by the exponentiation of graviton exchange in e.g. the eikonal approximation~\cite{Kabat:1992tb}.  In any case, once
a potential and Hamiltonian are constructed to reproduce the scattering amplitude, they can be applied more generally
to classical physics.

Having reviewed the overall kinematics of a scattering process that
captures its classical limit, we proceed to detail
the kinematics of the exchanged gravitons. This identifies the parts
of loop amplitudes that contribute in the classical limit, thus allowing us
to discard from the outset the parts that have no classical
contributions. 
The main observation is that, in the classical regime in which the total momentum transferred $\bm q$ is small compared to external 
momenta, the momentum of each individual graviton should be of the same order. To identify the relevant contributions we consider an internal 
graviton line with four-momentum $\ell = (\omega, \bm \ell)$ and,  following the method of regions~\cite{Beneke:1997zp, Smirnov:2004ym}, we consider the possible scalings of its momentum components:
\begin{align}
{\rm hard}: \quad (\omega, \bm \ell) &\sim (m,m) \,, \nn\\
{\rm soft}: \quad (\omega, \bm \ell) &\sim  (|\bm q|,|\bm q|) \sim J^{-1}\,(m |\bm v|,m |\bm v|) \,, \nn\\
\qquad {\rm potential}: \quad  (\omega, \bm \ell) &\sim (|\bm q| |\bm v|, |\bm q|) \sim J^{-1}\,(m |\bm v|^2,m |\bm v|) \,, \nn \\
\qquad {\rm radiation} : \quad  (\omega, \bm \ell) &\sim (|\bm q| |\bm v|, |\bm q| |\bm v|) \sim J^{-1}\,(m |\bm v|^2, m |\bm v|^2)\,,
\label{eq:modes}
\end{align}
where we take as reference scale $m = m_1 + m_2$ (or the external momentum), and we use
Eq.~\eqref{eq:classical_limit} to arrive at the second set of scalings in the above equation. 
Gravitons with hard ${\cal O}(m)={\cal O}(|\bm p|)$ momenta lead to quantum-mechanical contributions
because their energy component is too large, causing the matter fields to fluctuate far off shell. 
Gravitons in the soft region mediate long-range interactions, because $|\ell|\sim|\bm q|\sim |\bm b|^{-1}$, so they can contribute to a 
classical potential.  We use the velocity $0\le |{\bm v}| \ll 1$ to separate the soft region into potential and radiation regions. 
For small velocities, the modes in the potential region are off shell and carry little energy so they mediate interactions that are almost instantaneous, which is the hallmark of a classical potential.
The gravitons in the radiation region can be on shell so they can be emitted in a scattering process. They can also be reabsorbed 
by the system and contribute to its effective potential. This is the origin of the so-called tail effect~\cite{Thorne:1980ru}.
The modes in \eqn{eq:modes} identify the dominant contribution from
each region to generic loop integrals. Each of them is computed by expanding each loop momentum about the
given scaling and then integrating over the {\it full} phase space using
dimensional regularization.  To reconstruct the complete integral one
simply sum over all the regions. The apparent overcount stemming from
the integration over all momenta after expansion in each region is
only superficial: expanding momenta in one region about another leads
to scaleless integrals which vanish in dimensional regularization.
For further detail on the method of regions we refer the reader to
Ref.~\cite{Smirnov:2004ym}.

The above considerations, together with the observation that graviton
loops are scaleless and thus vanish in the potential region, imply that the contributions
of potential-region gravitons to the classical
potential \eqref{potential} have the following features:
\begin{enumerate}
	\item In all contributing diagrams, before and after reduction to a basis, the two matter lines do not intersect.

	\item Contributions where both ends of a graviton propagator attach to the same matter line are dropped.
	
	\item Every independent loop has at least one matter line.
	
	\item Terms with too high a scaling in $q$ or $\ell$ are dropped because they are quantum contributions. Eq.~\eqref{potential} implies that at $L$ loops 
         a for a given diagram with $n_m$ matter propagators, $n_g$ graviton propagators we can drop terms with more than $n_m + 2n_g - 3 L - 2$ powers of loop momentum 
         in the numerator.

\end{enumerate}

\begin{figure}
\begin{center}
\includegraphics[scale=.50]{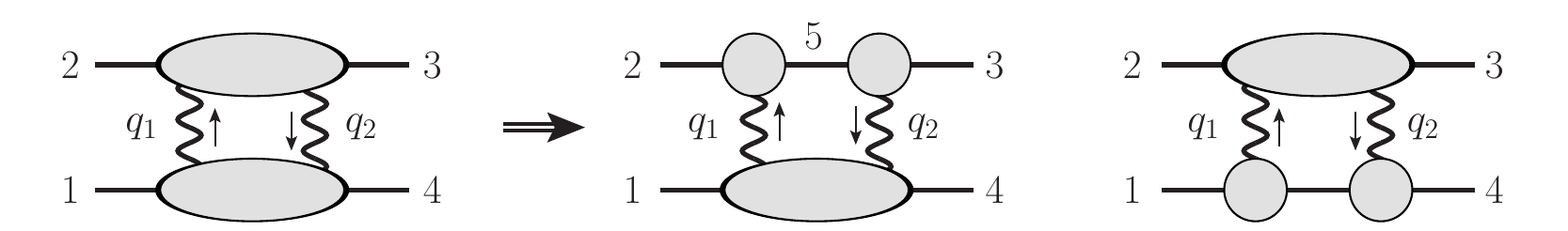}
\end{center}
\caption{Left-hand side: the part of the four-scalar one-loop amplitude that does not contain intersecting matter lines. Right-hand side: 
an identification of the part of the four-scalar one-loop amplitude that do not contain intersecting matter lines and have at least one matter line in the loop. Factorization of tree- and loop-level amplitude imply that the shaded blobs are tree-level amplitudes. }
\label{CutMatter}
\end{figure}

The first two of these features imply that the parts
of an $L$-loop amplitude that are relevant in the classical limit are
strictly a subset of the product of two two-scalar-$(L+1)$-graviton
tree amplitudes summed over the graviton states, together with scalar
propagators for each of the gravitons.
For example, at one loop this is a product of two gravitational
Compton amplitudes summed over the graviton states and divided by
$q_1^2q_2^2$ where $q_i$ are the momenta of the two gravitons. This is
shown graphically in the left-most diagram in Fig.~\ref{CutMatter}.

The third property of the contributions to the classical limit weeds
out part of the contributions appearing in the product of the two tree
amplitudes and keeps only those for which each independent loop has
at least one matter line before and after reduction to an integral basis.  One
starts with the terms having this property and in the process of
reducing to a basis of integrals keeps only those contributions that continue to
have this property. At one-loop level, the first step is shown on the
right-hand side of the arrow in Fig.~\ref{CutMatter}: there must be a
matter line in at least one of the two Compton amplitude factors.

While identified here from the perspective of the classical limit, the
contributions obtained this way have a natural interpretation in the
generalized unitarity method \cite{UnitarityMethod, Fusing,
TripleCuteeJets, BCFUnitarity, FiveLoop, BDKUniarityReview,
BernHuangReview, JJHenrikReview}, where they are referred to as
generalized cuts. The cut conditions---that is on-shell conditions for
the exposed lines---prevent those lines from being canceled in the
process of reduction to an integral basis. It is important to note that
the cut momenta are on shell only for the amplitudes represented by
the blobs; the propagators for the exposed lines are not placed on
shell in this procedure.

Factorization of tree amplitudes implies that the contributions given
by generalized cuts are expressed in terms of sums of products of tree
amplitudes; thus, one can directly apply the KLT relations to obtain
them in terms of amplitudes of scalars coupled to vector fields and
thus essentially use the KLT relations to obtain higher-loop
amplitudes.
As an example, the expression of the first cut on the right-hand side of \fig{CutMatter} is
\begin{align}
C_{\rm GR}^{\rm (a)} & = \sum_{h_1, h_2 }
M^\tree_3(3^s,q_2^{h_2}, -5^s)   \,
M^\tree_3(5^s , -q_1^{h_1}, 2^s) \,
M^\tree_4(1^s,q_1^{-h_1}, -q_2^{-h_1},4^s) \nn \\
& = \sum_{\lambda_1, \lambda_2, \tilde\lambda_1, \tilde\lambda_2} it \,  {\cal P}_{h_2} {\cal P}_{h_2} [A^\tree_3(3^s,q_2^{\lambda_2}, -5^s) \,
A^\tree_3(5^s , -q_1^{\lambda_1}, 2^s) \,
A^\tree_4(1^s,q_1^{-\lambda_1}, -q_2^{-\lambda_2},4^s)]  \nonumber\\
& \hskip 2 cm \null \times
 [A^\tree_3(3^s,q_2^{\tilde\lambda_2}, -5^s)   \,
  A^\tree_3(5^s , -q_1^{\tilde\lambda_1}, 2^s) \,
   A^\tree_4(4^s,q_1^{-\tilde\lambda_1}, -q_2^{-\tilde\lambda_2},1^s)] \,,
\label{cut}   
\end{align}
where $h_{1,2}$ label the physical states of the graviton,  $\lambda_{1,2}$ and $\tilde\lambda_{1,2}$ label the physical states of the corresponding gluons, ${\cal P}_{h_1,h_2}$ are projectors restricting the product of gluon states to be a graviton state (i.e. they project out the dilaton and the antisymmetric tensor) and we use the four-point  BCJ amplitude relation~\cite{BCJ}  to simplify the expression.  
Thus, the gravity generalized cut is expressed directly in terms of the components of gauge-theory generalized cuts.
In four dimensions, where physical states are labeled by their
helicity, the projectors ${\cal P}_{h_1,h_2}$ simply correlate the
helicities of the gluons, $\lambda_1=\tilde\lambda_1$ and
$\lambda_2=\tilde\lambda_2$, and the gravity cut is expressed in terms
of the four helicity configurations of the gauge-theory cut. In this
way, through use of the double copy, the basic building blocks are
non-abelian gauge-theory tree amplitudes.

The generalized unitarity method also provides an algorithm for
assembling the various contributions obeying the properties described
above while ensuring that terms that appear in several generalized
cuts are counted only once.
For reviews of the generalized unitarity method see
Refs.~\cite{BDKUniarityReview,BernHuangReview,JJHenrikReview} and in
the context of the classical limit of scattering amplitudes see 
Ref.~\cite{3PMLong}.

\subsection{Classical potential and classical observables from classical amplitudes}

Assuming that amplitudes evaluated in the classical limit are known, the next task is to find a generating function of classical observables whose form is \eqref{potential}. This generating function is understood as part of the Wilsonian-type effective action generated by integrating out graviton configurations that contribute to conservative physics \cite{Bern:2021yeh}. They may be potential-region gravitons \cite{CheungPM, 3PM, 3PMLong, Bern:2021dqo} or a mixture of potential and radiation region gravitons \cite{Bern:2021yeh}.
Constructing amplitudes from this effective action reveals that they exhibit classical parts, which scale in the large angular momentum limit as described in the previous section, and also ``super-classical'' parts, which dominate in the large angular momentum limit over the classical ones. Thus, the task is to consistently separate the classical part. Several methods have been proposed in this direction and we briefly summarize them here in no particular order. 

\vskip .3 cm 

{\bf 1.} Construct an effective two-body potential \cite{CheungPM,
3PM, 3PMLong}, which is then used in Hamilton's equations to generate
classical observables. If the Hamiltonian is independent of the
classical trajectory, as it is the case for the potential-graviton
contributions, a change in boundary conditions suffices to relate open
trajectory and bound orbit motion.

The effective two-body potential is obtained through by a matching calculation in which one demands \cite{CheungPM} that the scattering amplitudes of gravitationally-coupled scalars due to potential or mixed but time-symmetric gravitons are reproduced by an action containing only the positive-energy modes of the matter fields and with instantaneous (or energy- or time-independent) interactions
\begin{align}
H = A^\dagger \left(i\partial_t + \sqrt{\bm p^2+m_1^2}\right) A + B^\dagger \left(i\partial_t + \sqrt{\bm p^2+m_1^2}\right) B 
+ V(\bm p) A^\dagger A B^\dagger B \ ,
\label{NRaction}
\end{align}
with $V$ in Eq.~\eqref{potential}. The amplitudes following from this
action are matched order by order in Newton's constant with those of
the GR coupled to scalar fields of masses $m_1$ and $m_2$; at each
order one more coefficient of $V$ is determined: tree-level matching
fixes $c_1(\bm p)$, one-loop matching fixes $c_2(\bm p)$, etc. 
At a loop order $L$, with stronger-than-classical scaling at large
angular momenta are completely determined by the Hamiltonian coefficients determined through $(L-1)$-loop
order. For this reason they contain no new information and are referred to as ``iteration terms''.

We note that this effective potential can be systematically extended
to include quantum effects, see e.g. Refs.~\cite{Bjerrum-Bohr:2014zsa}; to this end one systematically keeps in 
the full-theory amplitude the desired quantum-suppressed terms. In particular, one may
include terms subleading in the large angular momentum expansion such
as graviton loops which would probe quantum gravity effects but one should
not include diagrams with intersecting matter lines, as they do not
contribute to long-range interactions.

\vskip .3cm 
{\bf 2.} Other amplitudes-based approaches construct a generating
function of open-orbit observables---the radial action---directly from
amplitudes or evaluate open-orbit observables in terms of matrix
elements of operators in the final state of the process.

The relation between the all-orders amplitude and the radial action
builds on the observation that the solution to the unitarity
constraint for an elastic two-particle S matrix is a phase.  Inspired
by the eikonal approximation \cite{Amati:1990xe, Saotome:2012vy,
Akhoury:2013yua, DiVecchia:2019myk, DiVecchia:2019kta,Bern:2020gjj},
the ``amplitude-radial action'' relation is \cite{Bern:2021dqo, Bern:2021yeh} 
\begin{align}
	i {\cal M}(\bm q)& = \int_{J}\,\left( e^{i I_r(J)} - 1 \right) ~ , \quad {\tilde I}_r(\bm q)=4E|\bm p|  \int d^{D-2}\bm b \, \mu^{-2\epsilon} e^{i\bm q\cdot \bm b I_r(J)} \nonumber \ ,
	\\
	{\tilde I}_r(\bm q)&=\frac{G}{|\bm q|^2}a_1(\bm p)+ \frac{1}{|\bm q|^3} \sum_{n\ge 2} {(G |\bm q|)^n (\ln \bm q^2 )^{n \text{ mod 2}}  a_n(\bm p)} \ ,
\label{eq:action}
\end{align} 
where $E$ is the total energy, $\bm b$ is the impact parameter and $\mu$ is the scale of dimensional regularization.
Classical observables are subsequently constructed through
thermodynamic-type relations (known for closed-orbit motion as the
first law of binary mechanics \cite{LeTiec:2015kgg}), e.g.
\begin{equation}
dI_r = \frac{\theta}{2\pi} dJ + \tau dE + \sum_a \langle z_a\rangle dm_a  \ ,
\label{dIr}
\end{equation}
where $\theta$ is the scattering angle, $\tau$ is the time delay and
$\langle z\rangle$ is the averaged redshift. This has been used to
systematically bypass iterated
contributions~\cite{Brandhuber:2021eyq, Kol:2021jjc,Bjerrum-Bohr:2021wwt}. 
The formalism of Ref.~\cite{Brandhuber:2021eyq} makes use of a heavy mass
version of the double copy~\cite{Brandhuber:2021kpo} to produce 
compact expression for the amplitude.
We refer the reader to the various original references for details.

The Kosower, Maybee, O'Connell (KMOC) formalism~\cite{Kosower:2018adc} constructs observables directly from amplitudes and
their cuts, dressed with the appropriate operators.  They are computed
as the difference between the expectation values of these operators in
the final and initial states,
\be
\Delta {\cal O} = \langle f |{\cal O}| f \rangle - \langle i |{\cal O}| i \rangle
\ee 
and the final and initial states are related by the S-matrix operator, 
\be
|f\rangle = S |i\rangle \ .
\ee
For example, the scattering angle is obtained from the change in momentum of matter particles. This approach will be summarized 
in Chapter 14 of this review~\cite{Kosower:2022yvp}.

To illustrate the methods let us now evaluate the ${\cal O}(G)$  and ${\cal O}(G^2)$ amplitudes in the classical limit and use them to find the effective potential and radial action.

\subsection{1PM}

The tree-level amplitude of two distinct massive scalars in the classical limit
due to graviton exchange is simple-enough to be obtained through a
Feynman graph calculation. It can also be obtained as a double copy of
two massive scalar amplitude due to gluon exchange. In this second
approach it is necessary to project out the dilaton-axion which is
part of the double copy of two vectors and couples to massive
particles.
This can be done while also focusing the long-range interactions
captured by this amplitude by evaluating only the pole part of the
amplitude,\footnote{We take this approach because constructing the complete 
four-point amplitude through the double copy requires subtracting out the dilaton 
exchange, which is present when the external particles are massive.}
\begin{align}
\label{polepart}
{\cal M}_4^\tree(1, 2, 3, 4)\bigr|^\text{long range} &= \frac{i}{q^2}\sum_h {\cal M}_3^\tree(1, 4, q^h){\cal M}_3^\tree(2, 3, -q^{-h}) 
\\
&= i^2\left(\frac{\kappa}{2}\right)^2 \frac{i}{q^2}\sum_{\lambda, {\tilde \lambda}} {\cal P}_h \,
 A_3^\tree(1, 4, q^{\lambda}) A_3^\tree(2, 3, -q^{-\lambda}) \nn \\
& \hskip 2.5 cm \null \times
 A_3^\tree(1, 4, q^{\tilde\lambda}) A_3^\tree(2, 3, -q^{-\tilde\lambda}) \ ,
\nonumber
\end{align}
where  ${\cal M}_3^\tree(i, j, q^h)$ are two-scalar-graviton amplitudes, $h$ tags the physical states of the graviton, ${\cal P}_h$ explained below is the projector removing the dilaton, and we 
used the double-copy form \eqref{threepointdoublecopy} of ${\cal M}_3^\tree(i, j, q^h)$. 
Particles with momenta $p_1$ and $p_4$ have mass $m_1$,  those with momenta $p_2$ and $p_3$ have mass $m_2$ 
and the sum runs over the physical states of the exchanged graviton. The sum over the physical polarizations of the graviton 
gives the physical-state projector,
\begin{equation}
\sum_{h}\pol(k)^{\mu\nu}_h\pol(-k)^{\alpha\beta}_{-h}
= \frac{1}{2}\mathcal{P}^{\mu\alpha}\mathcal{P}^{\nu\beta}
+ \frac{1}{2}\mathcal{P}^{\nu\alpha}\mathcal{P}^{\mu\beta}
- \frac{1}{D-2}\mathcal{P}^{\mu\nu}\mathcal{P}^{\alpha\beta} \,,
\label{CompletenessRelationGravity}
\end{equation}
where 
\begin{equation}
\mathcal{P}^{\mu\nu}(k)=\eta^{\mu\nu}-\frac{r^\mu k^\nu+r^\nu k^\mu}{r\cdot k} \, ,
\label{PhysicalStateProjector}
\end{equation}
and $r^\mu$ is an arbitrary null reference vector. Gauge invariance of the three-point amplitudes \eqref{polepart} guarantees that the reference vector drops out, so we can effectively take $\mathcal{P}^{\mu\nu}(k)\rightarrow \eta^{\mu\nu}$ and the physical-state sum \eqref{CompletenessRelationGravity} to be the numerator of the graviton propagator in de Donder gauge.  

The two three-point amplitudes can be obtained as double-copies of the two-scalar-gluon amplitudes, as in Eq.~\eqref{doublecopy3}.  They are
\be
i{\cal M}(1, 4, q^h) = \frac{\kappa}{2}{\cal P}_h (\sqrt{2}\varepsilon^\lambda_{\mu}(q) p_1^\mu)(\sqrt{2}\varepsilon^\lambda_{\nu}(q) p_1^\nu) =  \frac{\kappa}{2}  (2\varepsilon^h_{\mu\nu}(q) p_1^\mu p_1^\nu)  \ ,
\ee
where, as before, $h$ labels the physical states of the graviton and $\varepsilon^h_{\mu\nu}$ is transverse.
This defines the operator ${\cal P}_h$ used in Eq. (\ref{polepart}). 
Using this together with \eqref{CompletenessRelationGravity}, Eq.~\eqref{polepart} then becomes
\be
{\cal M}^\text{tree, class} = -\frac{16\pi i G m_1^2 m_2^2 }{q^2} (2\sigma^2-1) \ ,
\label{treeGW}
\ee
where $m=m_1+m_2$, $\nu=m_1m_2/(m_1+m_2)^2$, and $\sigma = p_1\cdot p_2/(m_1m_2)$.  Accounting for the nonrelativistic normalization of the amplitudes following from the action \eqref{NRaction}, the resulting 
${\cal O}(G)$ potential coefficient is
\be
c_1(\bm p) = \frac{M^4\nu^2}{E_1 E_2}(1 - 2\sigma^2) \ ,
\ee
where $M=m_1+m_2$, $\nu = m_1 m_2/M^2$ and $E_{1,2}=\sqrt{ {\bm p}^2+m_{1,2}^2}$ are the energies of the two incoming particles.

Similarly, comparing Eq.~\eqref{treeGW} with eq.~\eqref{eq:action} and using Refs.~\cite{Bern:2021dqo, Bern:2021yeh} it follows that the leading term of  the radial action is
\be
a_1(\bm p) = 16\pi M^4\nu^2 (2\sigma^2 -1) \ .
\label{treeRad}
\ee
Fourier-transforming to impact-parameter space\footnote{Note that this is a two-dimensional Fourier-transform, because the on-shell conditions on the external states constrain the momentum transfer $q$ to be two-dimensional.} leads, through Eq.~\eqref{dIr}, to the same scattering angle as the Hamiltonian.

\subsection{2PM}

The next contribution to the potential comes from the four-scalar one-loop amplitude. As we discussed, the generalized unitarity method provides an algorithmic construction for this and higher-loop amplitudes, while simultaneously seamlessly singling out the parts exhibiting 
the features required of the classical limit and interfacing with the double copy to organize gravity calculations in terms of simpler 
gauge theory ones.
The one-loop amplitude however is sufficiently simple so we can construct it without making use of the details of the general approach 
while still avoiding explicit use of Feynman diagrammatics.  

As we discussed on general grounds in Sec.~\ref{MatterGravitonKinematics}, to focus on the parts of the amplitude that do not contain intersecting matter lines it suffices to set to zero in the numerator of all contributing diagrams the squared momenta of the gravitons connecting the two matter lines -- momenta $q_1$ and $q_2$ on the left-hand side of Fig.~\ref{CutMatter}. 
Up to the overall factor of the two graviton propagators, this is the residue of the one-loop amplitude corresponding to the pole $q_1^2=0=q_2^2$, implying that
\be
{\cal M}_4^\text{1-loop} = \int \frac{d^d q_1}{(2 \pi)^d} 
\frac{i}{q_1^2}\frac{i}{q_2^2}\sum_{h_1, h_2} {\cal M}_4^\text{tree}(1, 4, q_1^{h_1}, -q_2^{h_2})
{\cal M}_4^\text{tree}(2, 3, -q_1^{h_1}, q_2^{h_2})+ \ldots \, ,
\label{oneloop}
\ee
where ${\cal M}_4^\text{tree}$ are gravitational Compton amplitudes and the ellipses represent terms that are not long-range classical.
This avoids discussing the details of assembling the two finer
contributions to the classical amplitude shown on the right-hand side
of Fig.~\ref{CutMatter} since \eqn{oneloop} automatically contains
both. At higher loops however the most efficient strategy is to use
the generalized unitarity method based on tree amplitudes with the
fewest numbers of legs.

The two Compton amplitude factors follow from the double copy of the dimensional reduction of higher-dimensional four-gluon amplitude, with two gluons taken in the extra dimensions. The dilaton-axion scalar is projected out from the product of each pair of intermediate gluons so the remainder are only the physical states of two gravitons.
The sum over each them gives the physical-state projector \eqref{CompletenessRelationGravity} used in the tree-level computation. A judicious choice of polarization-stripped amplitudes \cite{KoemansCollado:2019ggb} leads to a manifest cancellation of the reference vector. Such choices, which can involve adding terms that vanish on-shell to allow amplitudes to manifestly obey Ward identities, have been shown to always be possible \cite{Kosmopoulos:2020pcd}.

The Compton amplitude may also be obtained through the KLT relation, as in Eq.~\eqref{KLT4}.\footnote{Unlike Eq.~\eqref{polepart},
the dilaton contribution to the four-point tree amplitudes entering ${\cal M}^\text{1-loop}$ is projected out by simply choosing the external (cut) lines to be gravitons.} 
In this case ${\cal M}^\text{1-loop} $ is written as
\begin{align}
\label{oneloopKLT} 
{\cal M}_4^\text{1-loop} &= (- i)^2\left(\frac{\kappa}{2}\right)^4\int \frac{d^d q_1}{(2 \pi)^d}
\frac{i}{q_1^2}\frac{i}{q_2^2}(q^2)^2\sum_{\lambda_1,\lambda_2, \tilde\lambda_1, \tilde\lambda_2 } 
P_{h_1}P_{h_2}
A_4^\text{tree}(1, 4, q_1^{\lambda_1}, -q_2^{\lambda_2}) 
\\
&\hskip 1.5 cm \times A_4^\text{tree}(2, 3, -q_1^{\lambda_1}, q_2^{\lambda_2})
A_4^\text{tree}(1, 4, -q_2^{\tilde\lambda_2}, q_1^{\tilde\lambda_1})
A_4^\text{tree}(2, 3, q_2^{\tilde\lambda_2},-q_1^{\tilde\lambda_1})
+ \ldots \, ,
\nonumber
\end{align}
where $P_{h_1}$ and $P_{h_2}$ project out the dilaton and antisymmetric tensor from the product of two gluon states and the two factors of $q^2$ come from the four-point KLT relation. The sum over the 
gluon states is given by Eq.~\eqref{PhysicalStateProjector} and together with $P_{h_1}$ and $P_{h_2}$ gives again Eq.~\eqref{CompletenessRelationGravity}. 
In four dimensions and in spinor-helicity notation this is
straightforward~\cite{3PMLong}: one simply correlates the helicities
of the gluons in the two amplitude factors,
$(\lambda_1, \tilde\lambda_1), (\lambda_2, \tilde\lambda_2) \in \{
(+,+), (-,-)\}$, so the scalar states $\{ (+,-), (-,+)\}$ never appear
in the product. Four-dimensional methods continue to produce correct
results at ${\cal O}(G^3)$ \cite{3PMLong}; at higher orders however
more caution is necessary because of subtleties with dimensional
regularization~\cite{Bern:2021yeh}.

The result of either of these methods is then reduced to the standard one-loop basis of scalar box, triangle and bubble integrals; during the calculation we enforce the four requirements that weed out quantum contributions. Discarded contributions are diagrams with crossing matter lines and graviton loops and the only surviving ones are the box and the triangle integrals~\cite{BjerrumClassical, CheungPM} 
\be
\frac{{\cal M}^\text{1 loop}}{64 \pi^2 G^2 m_1 m_2} = 4m_1^3m_2^3(2\sigma^2-1)^2 (I_\text{Box} +I_\text{XBox})
- 3m_1m_2(5\sigma^2-1)( m_1^2 I_{\bigtriangleup}+m_2^2I_{\bigtriangledown})+\ldots ,
\label{1loopAmp}
\ee
where the ellipsis stand for terms that are not long-range or classical or both, and
\begin{align}
I_\text{Box}&= \int \frac{d^d\ell}{(2\pi)^2} \frac{1}{\ell^2 (\ell+q)^2 ((\ell+p_1)^2+m_1^2)((\ell-p_2)^2+m_2^2)} \,.
\end{align}
$I_\text{XBox}$ is obtained by interchanging $p_2$ and $p_3$ and $I_{\bigtriangleup}$ and $I_{\bigtriangledown}$  are obtained by 
removing one of the matter propagators with masses $m_2$ and $m_1$, respectively. While at this order integration is quite straightforward, 
it becomes less so at two loops and beyond; see Chapter 4 of this review~\cite{Blumlein:2022zkr} for modern techniques and results. 

Accounting for the nonrelativistic normalization of the amplitudes following from the action \eqref{NRaction}, the resulting 
${\cal O}(G^2)$ potential coefficient is
\be
c_2(\bm p) =\frac{M^5\nu^2}{E_1E_2} \left(
\frac{3}{4} (1-5\sigma^2)
- \frac{4M\nu E}{E_1E_2}\sigma(1-2\sigma)
- \frac{M^3\nu^2 E}{2\, E_1^2E_2^2}\left(1-\frac{E_1E_2}{E^2}\right)(1-2\sigma)^2
\right) ,
\ee
where $E=E_1+E_2$. One may recognize the first term in parenthesis as the coefficient of the triangle integrals in Eq.~\eqref{1loopAmp}; 
the other two terms originate from the the subtraction of the term with stronger-than-classical scaling present in the box integral.

Separating the iteration of the tree-level radial action \eqref{treeRad} as in Ref.~\cite{Bern:2021dqo}, leads to the ${\cal O}(G^2)$ term 
of the radial action is
\be
a_2(\bm p) = 6\pi^2 \nu^2 M^5 (5\sigma^2 - 1) \,.
\ee
As at ${\cal O}(G)$, observables following from the radial action thus derived agree with those following from the two-body Hamiltonian.

\subsection{Remarks and Outlook}

The methods summarized above have been used to derive the two-body potential and the radial action that capture  
the suitably-defined \cite{Bern:2021yeh} conservative open-orbit dynamics through ${\cal O}(G^4)$. 
An essential ingredient in these calculations has been the double-copy form of tree-level gravity amplitudes in terms 
of gauge-theory amplitudes.  Similarly, the KMOC formalism together with the double copy as a means for deriving the 
necessary amplitudes has been used to derive the impulse and energy loss 
through ${\cal O}(G^3)$~\cite{Parra-Martinez:2020dzs, Herrmann:2021lqe, Herrmann:2021tct}.   
Further progress may build on double-copy constructions with gauge-invariant kinematic numerators \cite{Brandhuber:2021eyq, Brandhuber:2021kpo} 
obtained from recent developments in the kinematic algebra of gauge theories~\cite{Chen:2019ywi, Johansson:2019dnu}.
Spin can also be incorporated
into this framework \cite{Bern:2020buy, Kosmopoulos:2021zoq}.  Here the double-copy properties are less obvious, though 
at least to quadratic order in the spins the gravitational Compton amplitudes have simple double-copy relations
to gauge theory, and so does the tree-level energy momentum tensor for any power of spin~\cite{Bern:2020buy}. 
A double copy for massive particles with spin including quantum effects was also discussed in Ref.~\cite{Johansson:2019dnu}.

\section{Conclusions}
\label{ConclusionSection}

In this mini-review, we  summarized the status of color-kinematics duality
and the associated double-copy construction,
including the basics of color-kinematics duality, the web of theories linked by the
double copy, and applications to gravitational-wave physics.  In recent
years there has been considerable interest in color-kinematics duality and the
associated double copy, especially towards finding new classical
solutions where the double copy holds (see e.g.
Refs.~\cite{Saotome:2012vy, Monteiro2014cda, Luna2015paa,
Ridgway2015fdl, Luna2016due, White2016jzc, Cardoso2016amd,
Goldberger2016iau, Luna2016hge, Goldberger2017frp, Adamo2017nia,
DeSmet2017rve, BahjatAbbas2017htu, CarrilloGonzalez2017iyj,
Goldberger2017ogt, Li2018qap, Ilderton:2018lsf, Lee:2018gxc,
Plefka:2018dpa, ShenWorldLine, Berman:2018hwd, Gurses:2018ckx,
Adamo:2018mpq, Bahjat-Abbas:2018vgo, Luna:2018dpt, Farrow:2018yni,
CarrilloGonzalez:2019gof, PV:2019uuv,Adamo:2020qru, Adamo:2021dfg}),
identifying supergravity theories admitting a double-copy construction (see
Tables~\ref{table-zoology-ungauged} and~\ref{table-zoology-gauged}),
and applying the double copy to physical problems such as precision
gravitational-wave computations~(see
e.g.~Refs.~\cite{3PM, 3PMLong, Bern:2021dqo, Brandhuber:2021eyq, Bern:2021yeh}).  
There has also been important progress on basic questions such as
identifying the underlying kinematic algebra behind  color-kinematics
duality~\cite{Monteiro2011pc, Monteiro:2013rya, Fu:2016plh,
Chen:2019ywi, Cheung:2017yef, Brandhuber:2021bsf, Ben-Shahar:2021zww}.

There are a number of obvious future directions which have attracted recent attention, seen exciting progress, and will be interesting to investigate further:
\begin{itemize}

\item Identifying new classes of classical solutions where the double copy holds,
  especially for cases that do no rely on the special properties of Kerr-Schild form of the
  metric~\cite{Monteiro2014cda, Luna2015paa, Ridgway2015fdl, BahjatAbbas2017htu, CarrilloGonzalez2017iyj}.  
   More generally, it would be important to find rules for
  choosing good coordinates and gauges that make double-copy relations
  more transparent.  

  \item  Realizing generalizations of scattering amplitudes in (A)dS that manifest the duality between color and kinematics~\cite{Albayrak:2020fyp,Armstrong:2020woi,Diwakar:2021juk,Cheung:2022pdk,Herderschee:2022ntr,Drummond:2022dxd}. 

\item Further understanding the underlying kinematic algebra behind the
duality between color and kinematics.  A natural expectation is that
the kinematic Jacobi identities are due to an infinite-dimensional Lie
algebra~\cite{Monteiro:2011pc}. Finding a complete description of such an algebra remains an open challenge, albeit with recent growing attention and progress~\cite{Cheung:2016prv,Chen:2019ywi,Chen:2021chy,  Ben-Shahar:2021doh, Cheung:2021zvb, Brandhuber:2021bsf,Ben-Shahar:2021zww}.

\item Expanding the web of theories linked by double-copy relations 
described in Sect.~\ref{ZoologySection1}.  This includes finding
further non-gravitational examples beyond those listed in
Table~\ref{table-zoology-nongrav} and understanding whether all
supergravity theories are necessarily double copies.

\item Carrying out new state-of-the-art computations of physical 
or theoretical interest. Recent examples are high-order calculations
in gravitational wave physics~\cite{3PMLong, Bern:2021dqo,
Bern:2021yeh}. The recent construction of the six-loop integrand of
$\mathcal N = 4$ super-Yang-Mills theory~\cite{Carrasco:2021otn} suggests
that analogous progress is possible for $\mathcal N = 8$ supergravity,
with a goal of obtaining the ultraviolet behavior.

\item Identifying and developing novel directions.  Recent examples include 
finding color-kinematics duality in a non-Abelian version of
Navier-Stokes equation of fluid mechanics~\cite{Cheung:2020djz},
Chern-Simons theory~\cite{Ben-Shahar:2021zww}, quantum
entanglement~\cite{Cheung:2020uts} and field-space geometry~\cite{Cheung:2022vnd,Cohen:2022uuw}.

\item Finding new connections between the double copy and other advances 
in scattering amplitudes, such as the
amplituhedron~\cite{Arkani-Hamed:2013jha, Trnka:2020dxl}, integrated
high-loop results for planar $\NeqFour$ super-Yang-Mills theory (see
e.g. Ref.~\cite{Dixon:2020bbt}).

\end{itemize}

The duality between color and kinematics and the
associated double-copy structure offer a novel perspective on gravity
theories compared to more traditional geometric approaches.  They were
originally formulated for flat-space perturbative scattering
amplitudes, where they offer insight and tools to address a variety of
problems. Based on large numbers of known examples, the double copy applies much
more generally, not only to classical solutions but also to a web of
interlocked gravitational and nongravitational theories.  The
surprisingly large web of theories included in \fig{FigWeb} suggests
that (quantum) field theories theories have new nontrivial hidden constraints, as
suggested by the fact that the number of building blocks is smaller than 
the number of consistent theories.  In the coming years, it will be fascinating to find out the reach of these
ideas both on the computational and theoretical sides.

\vskip .3 cm 

\section*{Acknowledgments}
We are particularly grateful to Oliver Schlotterer for detailed feedback on the manuscript. 
We also thank Keith Ellis for pointing out typographical errors in an earlier version of this review.
Z.B. is supported by the U.S. Department of Energy (DOE) under grant
no.~DE-SC0009937 and by the Mani L. Bhaumik Institute for Theoretical
Physics.  J.J.M.C. is grateful for the support of Northwestern
University, the DOE under contract DE-SC0021485, and by the Alfred P. Sloan 
Foundation. 
R.R.~is supported by the U.S. Department of Energy (DOE)
under grant no.~DE-SC00019066. 
The work of M.C. is supported by the Swedish Research Council under grant 2019-05283.
The research of M.C.~and~H.J. is also
supported by the Knut and Alice Wallenberg Foundation under grants KAW 2018.0116 ({\it From Scattering Amplitudes to Gravitational Waves}) and KAW 2018.0162, and the Ragnar S\"{o}derberg Foundation (Swedish Foundations' Starting Grant).

\vskip 1 cm 

\bibliography{chapter2}

\end{document}